\documentclass[runningheads]{llncs}
\usepackage{graphicx}
\usepackage[english]{babel}
\usepackage[autostyle, english = american]{csquotes}
\MakeOuterQuote{"}
\usepackage[table]{xcolor}

\usepackage{multirow}
\usepackage{float}
\usepackage{nameref}
\usepackage{indentfirst}
\usepackage{url}
\usepackage{hyperref}
\usepackage{nameref}

\begin{document}
\title{How to explain it to data scientists?}
\subtitle{A mixed-methods user study about explainable AI, using mental models for explanations}
\titlerunning{How to explain it to data scientists?}
%
\author{Helmut Degen\inst{1}\orcidID{0000-0002-5200-8165}
\and
Ziran Min\inst{1}\orcidID{0009-0006-2820-754X} 
\and
Parinitha Nagaraja\inst{1}\orcidID{0009-0000-3900-8163}
}
\authorrunning{H. Degen et al.}
%
\institute{Siemens Corporation, Foundational Technologies, 755 College Road East, Princeton NJ 08540, USA \\ 
\email{\{helmut.degen,ziran.min,parinitha.nagaraja\}@siemens.com}
}
\maketitle 
%
\begin{abstract}

In the context of explainable artificial intelligence (XAI), limited research has identified role-specific explanation needs. This study investigates the explanation needs of data scientists, who are responsible for training, testing, deploying, and maintaining machine learning (ML) models in AI systems. The research aims to determine specific explanation content of data scientists. A task analysis identified user goals and proactive user tasks. Using explanation questions, task-specific explanation needs and content were identified. From these individual explanations, we developed a mental model for explanations, which was validated and revised through a qualitative study (n=12). In a second quantitative study (n=12), we examined which explanation intents (reason, comparison, accuracy, prediction, trust) require which type of explanation content from the mental model. The findings are: F1: Explanation content for data scientists comes from the application domain, system domain, and AI domain. F2: Explanation content can be complex and should be organized sequentially and/or in hierarchies (novelty claim). F3: Explanation content includes context, inputs, evidence, attributes, ranked list, interim results, efficacy principle, and input/output relationships (novelty claim). F4: Explanation content should be organized as a causal story. F5: Standardized explanation questions ensure complete coverage of explanation needs (novelty claim). F6: Refining mental models for explanations increases significantly its quality (novelty claim).

\end{abstract}


\section{Introduction}
\label{Introduction}

Artificial intelligence (AI) is used in many consumer and industrial applications. Many challenges remain in making AI systems \footnote{In the context of this paper, AI systems refer to systems that use technologies including machine learning, deep learning, and/or large language models.} ethical, fair, and human-centered \cite{Garibay2023-article}.  One of them is the need for explanations. This research is looking into explainable AI (XAI) for a futuristic system that partially automates the tasks of a data scientist in generating, deploying, and maintaining ML models. 

One important characteristic of AI technology is the possibility that an identified outcome can be incorrect from a human perspective. In this paper, we refer to this phenomenon as the "uncertainty premise." There are several underlying reasons for the uncertainty premise: (1) the use of statistical methods to determine outcomes \cite{MacKay2003-book-learning,Murphy2012-book-learning,Aggarwal2013-book-learning,Chapelle2010-book-learning}, (2) the opacity of black box models \cite{Carvalho2019machine-Article,Goodfellow2016-Book}, (3) the inherent non-deterministic behavior of many AI systems \cite{Lecun1998aneural-article,Goodfellow2016-Book}, (4) the use of non-representative datasets \cite{Kotsiantis2006ImbalancedData-article,Chawla2004specialImbalancedData-article,He2009learningImbalancedData-article}, and (5) hallucinations \cite{Barman2024-Hacculination-article,Huang2023-Surveyhallucinationlargelanguage-arxiv,Ji2023survey-article}.

The uncertainty premise is particularly a challenge for industrial and safety-critical systems \cite{Saraf2020-inproceedings}. To address uncertainties, we apply an explainable AI approach for AI systems "that can explain their rationale to a human user, characterize their strengths and weaknesses, and convey an understanding of how they will behave in the future" \cite[p. 44]{Gunning2019-article}. We want to stress here that an (uncertainty addressing) explanation is content provided to the user \textit{in addition} regarding the outcome of an AI system.

The application domain in this research study is industrial applications, including manufacturing, building technologies, mobility, and healthcare. 


To derive the research questions and metrics, the Goal/Metric/Questionnaire (GQM) approach is employed \cite{Basili1992-techreport}. 

The main research goal is to understand the explanation needs of the user role "Data Scientist" (RQ 1). Another research question is to understand which specific explanation content supports different explanation intents (RQ 2). Finally, we wanted to know which explanation intents provide value to data scientists and which do not (RQ 3).

To achieve the research goal, the following research questions are formulated:
\textbf{RQ 1:} Which explanation content does a data scientist need for selected goals and selected tasks?\\
\textbf{RQ 2:} Which explanation content does a data scientist help with for different explanation intents?\\
\textbf{RQ 3:} Which explanation intent has which value for data scientists?\\

\begin{figure}[!h]
	\includegraphics[width=1.0\textwidth,angle=0]{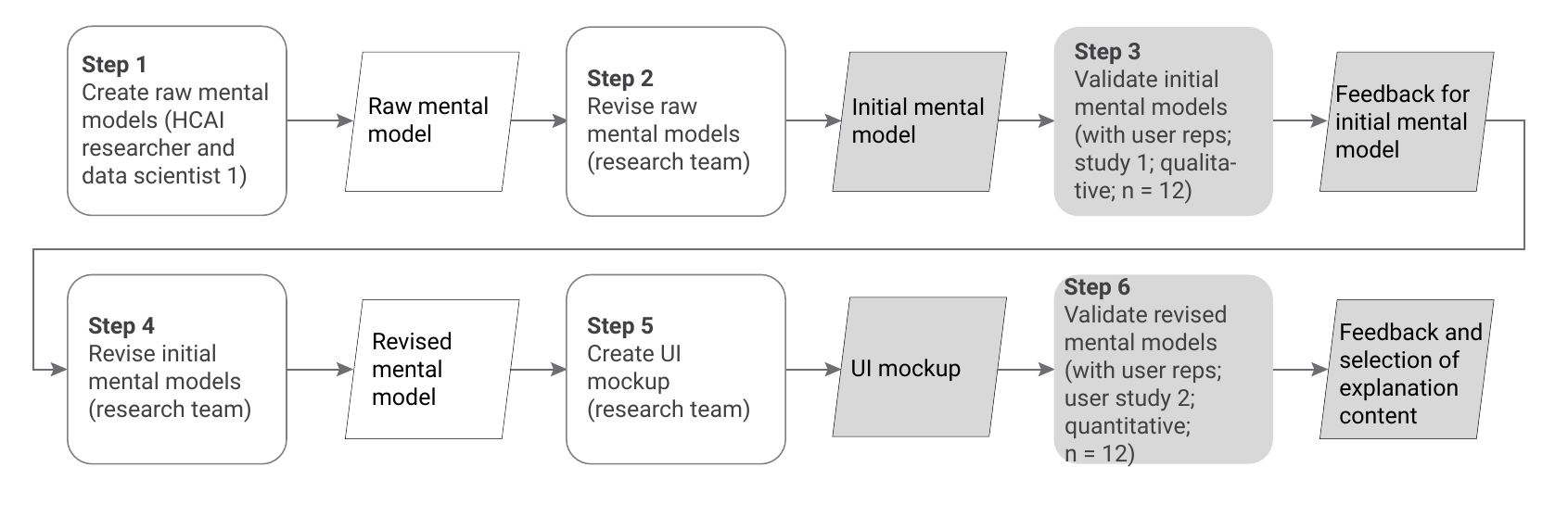}
	\caption{Research approach; step 3 and step 6 with the results are reported in this paper}
	\label{fig:ResearchApproach}
\end{figure}

To answer the research questions, we deployed a mixed-method approach \cite{Clark2016-Book}.  The research team consisted of three people: two data scientists (representatives of the target user group) and one human-centered AI (HCAI) researcher. The chosen approach is an evolution of previously deployed approaches \cite{Degen2023-inproceedings,Degen2024-inproceedings} with the use of mental models \cite{Johnson-Laird1986-book}. The mental model for explanations was developed in six steps (see Figure~\ref{fig:ResearchApproach}). Step 3 and step 6 are reported in this paper. 

In previous studies \cite{Degen2023-inproceedings,Degen2024-inproceedings}, steps 2, 3, and 4 were not performed. These steps were introduced to increase the quality of the mental model for explanations. In this paper, we report the results of steps 3 and 6 (including their inputs). The results of step 3 answer RQ 1 and the results of step 6 answer RQ 2 and RQ 3.

The contribution of this study is multi-fold: it explores a new methodology to use mental models for explanations that are user role-specific and task-specific. In addition, the revised mental model of explanations identifies new properties of explanations that can be added to known explanation properties and content types. 

In Section \ref{sec:RelatedWork} of the paper, we discuss the related work. Section \ref{sec:ApplicationDomain} introduces the futuristic intelligent system for the generation, deployment, and maintenance of ML-models and the user role "Data Scientist" with its user goals and user tasks, related to the futuristic intelligent system. Section \ref{sec:Study1} describes the design of the qualitative study with its results. In section \ref{sec:Study2}, the second, quantitative user study is described with its results. Section \ref{sec:Findings} provides a discussion of the results, the key findings, limitations of the work, and outlines future work.


\section{Related work}
\label{sec:RelatedWork}

We include here related work that includes explanation types, data scientists (user group and user tasks), and the use of mental models to identify needed explanations.


\subsection{Explainability}

Many scholars address XAI related research questions (for literature surveys see for instance \cite{Adadi2018-article,Angelov2021-article,Dosilovic2018-inproceedings,Guidotti2019-article,Hu2021-proceedings,Islam2022-article,Mueller2019-article,Vilone2020-article}). Many research papers report the evaluation and/or comparison of different kinds of explanations to understand their influence on understandability, trustworthiness and other explainability qualities. The following types of uncertainty addressing explanations have been identified so far: Prediction scope: 1) Global and local \cite{Arrieta2020-article,Adadi2018-article,Hoffman2018-article}; 2) Model view: Black box and white box \cite{Rudin2019-article}; 3) Intended use: Justify, control, improve, discover \cite[p. 52142--52143]{Adadi2018-article}; trustworthiness, causality, transferability, informativeness, confidence, fairness, accessibility, interactivity, privacy awareness \cite[p. 8--10]{Arrieta2020-article}; understandability, predictability \cite[p. 8]{Mohseni2021-article}; actionability \cite{Degen2023-inproceedings}; 4) Explanation structure: Singular, "show me your work" \cite{Degen2023-inproceedings}; 5) Domain types: AI domain, application domain \cite{Degen2023-inproceedings}, system domain \cite{Degen2024-inproceedings}; 6) Outcome comparison: contrastive and counterfactual explanations \cite{Miller2019ExplanationIA,Stepin2021-contrastive_counterfactual_explanations-article}, confidence measures \cite{Waa2020_ConfidenceLevel_2020,Zhang2020_Confidence_2020}; 7) Time dimension: backward-looking, forward-looking \cite{Degen2024-inproceedings}.


\subsection{Derivation of mental model of explanations}


When reviewing XAI and mental model literature, the criteria for the review are to find studies that: C1) use mental models to capture uncertainty addressing explanations, C2) demonstrate how to elicit, validate, and improve mental models with target user group representatives, and C3) create and validate mental models before system design or implementation.

A mental model is an internal representation of external reality \cite{Johnson-Laird1986-book}, aiding understanding, reasoning, and prediction in a domain \cite{Gentner2001-incollection}. In explainability, it should encapsulate explanation content tailored to a user role, task, and goal \cite[P. 9]{Hoffman2019metrics-misc}. This research employs a mental model of explanations that includes outcomes, inputs, and context of use, not just the AI system.

\cite[p. 5]{Hoffman2023-MeasureXAI-FrontiersinCS-article} provides an overview of methods to create mental models, using the 'diagramming task' in mixed-method studies. \cite{Garcia2018-XAI-MentalModel-inproceedings} describes a 'speak-aloud' method where an expert rationalizes system behavior while watching videos, visualized in a diagram. This method requires a stimulus showing system behavior (not meeting C3). \cite{Merry2021-article} introduces a new XAI definition, using team members' mental models to align AI explanations. Effective explanations need 'audience, language, and purpose' \cite[p. 5]{Merry2021-article}, but it doesn't describe how to elicit or validate mental models (not meeting C2). \cite{Onari2023-XAI-MentalModel-arxiv-article} uses Fuzzy Cognitive Maps (FCM) \cite{Kosko1986-FuzzyCognitiveMaps-IJMMS-article} to represent medical experts' mental models. FCMs predict trust levels for ML models but don't identify needed explanation content (not meeting C1). \cite{Rutjes2019-XAI-MentalModel-CHI-inproceedings} emphasizes the importance of users' mental models for effective explanations but doesn't elaborate on elicitation or validation (not meeting C2). \cite{Sheridan2023-MentalModel-InProceedings} used design thinking workshops to create mental models but lacked target user group representatives (not meeting C2). \cite{Degen2023-inproceedings,Degen2024-inproceedings} describes a method to elicit mental models for explanations, meeting C1 and C2, but didn't improve the model based on user feedback (not meeting C2).

In conclusion, none of the discussed work uses a comparable technique to elicit and validate mental models for explanations intended as explanation content for AI systems.


\section{Application domain: Responsibilities of a data scientist}
\label{sec:ApplicationDomain}


\subsection{Intelligent system to generate, deploy, and maintain ML-models}
\label{sec:ApplicationDomain:IntelligentSystem}

When we designed the study, we considered a futuristic intelligent system capable of generating, deploying, and maintaining ML models semi-automatically with minimal involvement from data scientists. If an ML model violates defined quality criteria during operation, the intelligent system is capable of identifying responsive actions. A diagram of the futuristic system is depicted in Figure~\ref{fig:IntelligentSystem}.


In the remainder of this paper, the intelligent system for generating, deploying, and managing ML models is referred to as the \textit{system under design (SuD)}.


\subsection{User role: Data Scientist}
\label{sec:ApplicationDomain:UserRole}

Our target user role is a data scientist (DS). We performed the goals and task analysis with two individuals experienced in the role and responsibilities of a data scientist. When identifying goals and tasks, we assumed the existence of the SuD. We identified the following user goals (UG) and user tasks (UT): UG 1: Make model available that meets the defined objectives (Rank 1) and UG 2: Make model available within project constraints (e.g., time, budget, available resources) (Rank 2).

To achieve the two goals, a data scientist performs the following user tasks, using the SuD: UT 1: DS (re) defines objectives (incl. scope, technical objectives, metrics, requirements, ranking, project constraints); UT 2: DS selects combination (incl. ML-model, data, and model parameters); UT 3: DS maintains ML-model. The three user tasks support the two user goals.

Since UT 1 focuses on defining the objectives and entering them into the SuD, the task does not require uncertainty addressing explanations. User tasks UT 2 and UT 3 assume that the SuD identifies the task-specific outcome. Therefore, uncertainty addressing explanations are needed.

For all three user tasks, we have identified the intent, the inputs, and the outputs. For user tasks UT 2 and UT 3, we have answered the following explanations questions (EQs) to identify task specific explanation content: EQ 1: Which explanation content do you need to understand \underline{how} the outcome was determined? EQ 2: Which explanation content do you need to understand \underline{why} the outcome was determined? EQ 3: Which explanation content do you need to understand the \underline{validity} of the outcome? EQ 4: Which explanation content do you need to \underline{predict} the effectiveness of the outcome in future use?

For user tasks UT 2 and UT 3, the inputs, outputs, and answers to the explainability questions are documented in the Appendix~\ref{app:sub:UserTask}. The use of predefined explanation questions per user task is an evolution of the methodology to identify explanation content, compared to \cite{Degen2024-inproceedings}.


\section{Study 1: Qualitative study}
\label{sec:Study1}


\subsection{Study objective, participants, preparation}
\label{sec:Study1:Goal}

The objective of the study is to validate the initial mental model for explanations and identify areas for improvement.

We use seven screening criteria for the selection of participants (see Appendix~\ref{app:sub:Screener}). We planned to recruit participants from the Siemens organization. Participants qualify if they have at least three years of industrial experience with ML models and answer at least four out of six other screener questions with 'yes.' To reach saturation with a homogeneous study sample, the target sample size is set to twelve participants \cite[p. 7]{Hennink2022-article} \cite[p. 74]{Guest2006-article}.

Based on the results of the user goal, user tasks, and explanation analysis, we derived two initial mental models for explanations: one for user task 2 and one for user task 3 (see Appendix~\ref{app:sub:InitialMM}). The visualization of the mental model is a technique called 'diagramming task' \cite{Hoffman2023-MeasureXAI-FrontiersinCS-article}.


\subsubsection{Initial mental model for UT 2 Generate ML model}

The initial mental model for explanations for UT 2 (Generate ML model) contains the following explanation elements (see Figure~\ref{fig:Task1_InitialGenerateMLModel} in the Appendix).

\textit{E 1 Objectives and objective priorities} This explanation element contains metrics such as accuracy, precision, data similarity, data type, data volume, training time, F1 score, priorities, and others.

\textit{E 2 Combinations (testing view)} The explanation element E 2 contains three explanation elements: E 2.1 Combinations. Per combination, it contains data set, model, model parameters, quantitative information, training metrics (e.g., loss function, bias/variants function); E 2.2 Ranking approach; E 2.3: Ranked combinations. Per combination: data set, model, model parameters, quantitative information, testing metrics (e.g., loss function, bias/variants function), performance metrics (e.g., ROC curve, PR curve, confusion matrix).

\textit{E 3 Combinations (evaluation in production-like environment)} The explanation element E 3 contains two explanation elements: E 3.1 Ranking approach; E 3.2 Ranked combinations. Per combination, it contains data set, model, model parameters, quantitative information, performance, and cost metrics (e.g., inference time, preprocessing time, precision, variant function).

\textit{E 4 Combined combinations} The explanation element E 4 contains two explanation elements for combined combinations, including testing and project view: E 4.1 Ranking approach; E 4.2 Ranked combinations. Per combination, it contains data set, model, model parameters, and quantitative information.


\subsubsection{Initial mental model for UT 3 Maintain ML model}

The initial mental model for explanations for UT 3 Maintain ML model contains the following explanation elements (see Figure~\ref{fig:Task2_InitialMonitorMLModel} in the Appendix).

\textit{E 5 Root cause} This explanation element E 5 contains four explanation elements: E 5.1 Objective and objective priorities: Metrics, such as accuracy, precision, data similarity, data type, data volume, training time, f1 score, priorities, and others; E 5.2 Context and selected symptom: Real world data, operational results, model monitoring reports (incl. deviations / symptoms); E 5.3: Ranking approach; E 5.4: Ranked root causes for selected symptoms. Per root cause: Symptom, real world data, quantitative information.

\textit{E 6 Responsive action (based on historical data)} The explanation element E6 contains two explanation elements: E 6.1 Ranking approach; E 6.2 Ranked responsive actions for selected root cause (historical view). Per responsive action: root cause, quantitative information.

\textit{E 7 Responsive action (based on evaluation in production-like environment)} The explanation element E 7 contains two explanation elements: E 7.1 Ranking approach; E 7.2 Ranked responsive actions for selected root cause. Per responsive action: Root cause, quantitative information.

\textit{E 8 Combined responsive action (historical and projection view)} The explanation element E 8 contains two explanation elements: E 8.1 Ranking approach; E 8.2 Ranked responsive actions for selected root cause. Per responsive action: Root cause, quantitative information.


\subsection{Study protocol and data analysis}
\label{sec:Study1:Protocol}

To evaluate the initial mental model for explanations, a semi-structured interview with each participant was conducted. The study protocol is shown in the Appendix~\ref{app:sub:protocols}. The HCAI expert conducted the interview. The two domain experts participated in each interview and answered AI domain-specific questions that required in-depth knowledge of the AI domain.

The two domain experts and the HCAI expert ("research team") entered the answers to steps 5-7 and steps 10-12 individually and independently into a spreadsheet. The research team reviewed all the entered answers together in a workshop, applying open coding. Based on the discussion, the research team decided together if the mental model for explanations should be modified, and if so, where and how.


\subsection{Recruited participants}
\label{sec:Study1:RecruitedParticipants}

We recruited twelve participants from the Siemens organization. The recruitment took place through the personal network of the research team. All participants have between 3 and 36 years of industrial experience with ML systems. They have all been involved in data selection, model training, model testing, and model fine-tuning. Ten out of twelve have experience with model integration and deployment, and seven out of twelve have experience with model maintenance. Five participants are based in Germany, and seven are based in the United States. Ten participants were male, and two were female. The participants have various experiences with different ML tasks (see Table~\ref{tab:Participant1}).

\begin{table}[!h]
	\footnotesize
	\caption{Participants' experience with different model tasks (n = 12)}
	\label{tab:Participant1}
	\begin{tabular}{>{\centering\arraybackslash}m{2.3cm} >{\centering\arraybackslash}m{2.3cm} >{\centering\arraybackslash}m{2.3cm} >{\centering\arraybackslash}m{2.3cm} >{\centering\arraybackslash}m{2.3cm}} 
		\hline
		Classification & Segmentation & Time Series Forecasting & Natural language processing & Reinforcement learning\\
		\hline
		12 & 8 & 11 & 10 & 8\\
		\hline
	\end{tabular}
	Read example: Eight (out of twelve) participants have experiences with Segmentation.
\end{table}

The interviews took place from August 22 through December 9, 2024. All participants gave their explicit consent to conduct the interview and to use their anonymized data for publication. All interviews were conducted in English. Each interview lasted between 60 and 120 minutes. All interviews were conducted remotely using Microsoft Teams and were recorded. The participants did not receive a monetary incentive.


\subsection{Study result}
\label{sec:Study1:Results}

\subsubsection{Received feedback for mental model 1 (Generate ML-model)}
\label{sec:Study1:Feedback:E1}

\textit{E 1: Objectives} Most participants expressed expectations to extend box E 1 to include more than just objectives. The type of problem that the AI attempts to solve should be considered (P3). The input should consider a separation between constraints and objectives (P1, P3, P7, P8, P9). The objectives should distinguish between the development phase (i.e., training and testing the model) and the operation phase (after model deployment) (P4, P10). The objectives and constraints should be categorized into the application domain, system domain, and AI domain (P9). It should also be possible to identify benchmarks against which the new models will be assessed (P5). Additionally, it should be possible to define hardware and software requirements as well as model types and model architectures as constraints (P8, P11, P12). The SuD should allow for the entry of data sets (P12). An Operational Design Domain (ODD) should be added to the inputs (P10).

\textit{E 2: Combination (testing view)} For explanation box E 2, participants expected more explanation content. Some participants expected application domain evidence in box E 2 (P1, P2). Another feedback was to add metadata to the data set, including how balanced the data set is, as well as synthetic data (P3, P7, P10, P11). Data governance for the listed data set should also be added as an explanation (P8). A box to describe the raw data, pre-processing algorithms before a model is trained and tested, as well as data mining insights, is missing (P9, P12). The shown metrics, although meant to be examples, are incomplete. Some said robustness metrics (P9, P10) are missing. If differences exist between the testing view and the production-like environment view, those differences should be highlighted (P10). The ranking approach for ranked combinations should be accessible (P4).

\textit{E 3: Combination (evaluation in production-like environment)} The ranking approach for ranked combinations should be accessible (P4). In addition, the simulation cases should be accessible and should include out-of-distribution data (P7). The term "production-like environment" should be replaced by "simulated production environment" (P7).

\textit{E 4: Combined combination} For the combined combination, considering the testing view and the production-like environment view, there should be access to interim outcomes so that data scientists can prune model explorations that may not comply with the defined constraints and may not achieve the defined objectives (P8, P10). For the selected combination, its strengths and weaknesses should be summarized (P9). The ranking approach for ranked combinations should be accessible (P4). While the combinations are being explored, the data scientist should be able to see the interim or final results to prune combinations that look unpromising and to make adjustments to the inputs to pursue certain objectives (P11).


\subsubsection{Received feedback for mental model 2 (Maintain ML-model)}
\label{sec:Study1:Feedback:E2}

\textit{E 5: Root cause} The objectives box (E 1) from the previous diagram should be reused in this explanation model (P1). Symptom information should be included as an explanation. It should contain changes in the environment, application, system, or AI domain (P2, P10). The data should be expressed as quantitative and qualitative data (P2, P10). Furthermore, the symptom should have access to real-world data that provide evidence for the existence of a symptom (P1, P6). In case symptoms appear frequently and systematically, such symptoms should be aggregated as symptom patterns that can be specific to one device (e.g., F1 score of industrial PC A goes down every Monday morning from 6 am to 9 am) or for multiple devices (e.g., F1 scores of industrial PCs A, B, and C go down every Monday morning from 6 am to 9 am) (P9). The components impacted by a root cause should be identified (P10). The impact of a symptom and/or a root cause on objectives should be identified and quantified (P10).

\textit{E 6: Responsive action (based on historical data)} For each responsive action, it should be transparent which symptoms and root causes are addressed (P8). For each responsive action, the following information should be displayed: impacted system component, quantitative information, and impact on objectives (P10). The term "Historical data" should be replaced by "Historical view" to avoid a conflict with training or testing data (P10).

\textit{E 7: Responsive action (based on evaluation in production-like environment)} The Operational Design Description, including the simulation cases, should be accessible (P4). For each responsive action, the following information should be displayed: impacted system component, quantitative information, and impact on objectives (P10).

\textit{E 8: Combined responsive action (Historical and production-like environment)} For the selected responsive action, the cost for repair, the cost/benefit ratio, the time for repair, the expected downtime of the machine or system, and the operational cost for not taking action to address the root cause should be displayed (P8). The decisions and their explanations should be presented along a timeline (P8). The sequence of the information should be organized in a causal chain: context, symptom, root cause, responsive action (P12).


\subsection{Revised mental model for explanations}
\label{sec:Study1:ResultsMM}

After analyzing the feedback from user study 1, we revised the two mental models of explanations. The revised version for user task 1, "Generate ML model," is depicted in Figure~\ref{fig:RevisedMentalModel1}, and the revised version for user task 3 is depicted in Figure~\ref{fig:RevisedMentalModel2}.


\subsubsection{Revised mental model for UT 2 Generate ML Model}

\begin{figure}[!h]
	\includegraphics[width=1.0\textwidth]{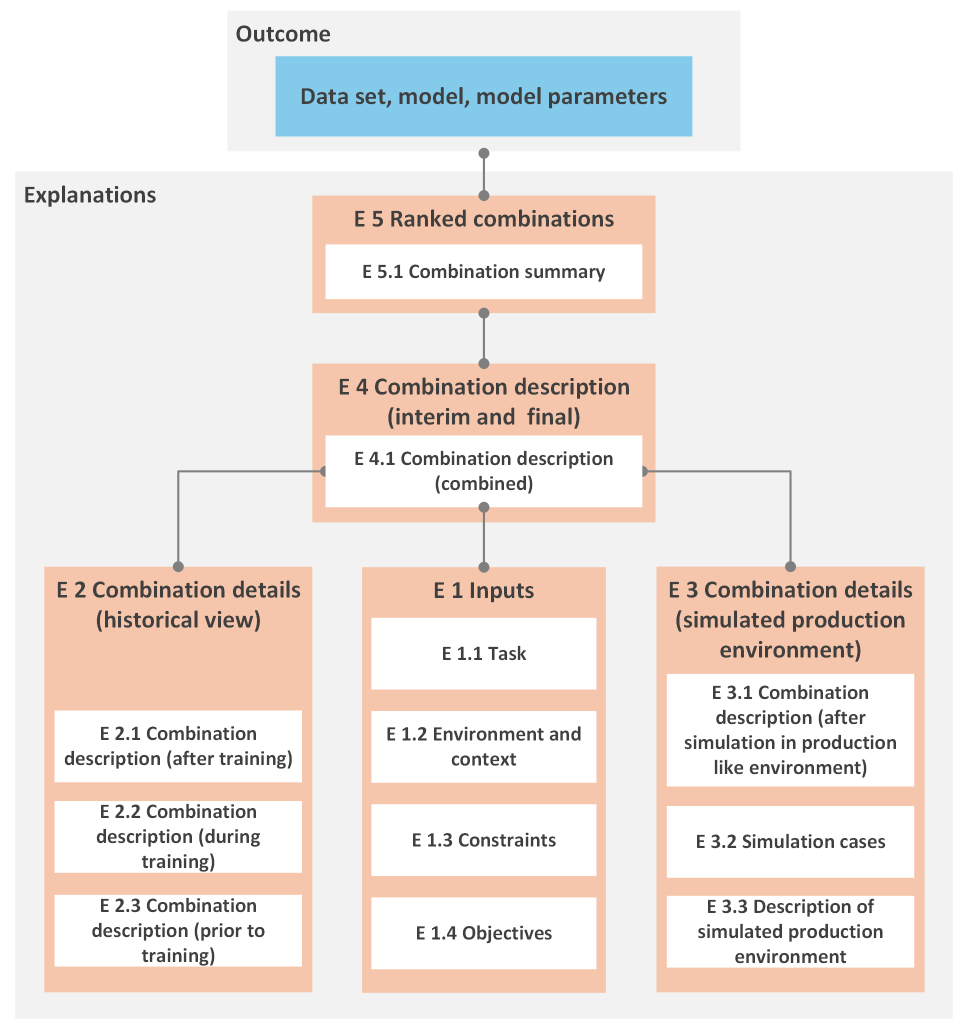}
	\caption{Revised mental model for user task "UT 1 Generate ML-model"}
	\label{fig:RevisedMentalModel1}
\end{figure}

The revised mental model for explanations for UT 2 (Generate ML model) contains the following explanation elements (see Figure~\ref{fig:RevisedMentalModel1}).

\textit{E 1 Inputs} The explanation element E1 contains the following explanation elements: E 1.1 Task: Application domain: Industry / vertical (e.g., PCB manufacturing); intended use (e.g., visual inspection to identify quality problems of PCBs; identify live cancer on x-ray images for adults); E 1.2 Environment and context:  Application domain: Environmental conditions (e.g., lighting conditions, dust, temperature range, humidity range, weather conditions, conveyor belt speed, camera, expected data drift); system context (e.g., network throughput, frequency of receiving data); E 1.3 Constraints: Development phase: system domain (e.g., available hardware, metrics), AI domain (e.g., model architecture, model types, technology, data set with meta-data, metrics); operation phase: application domain (environmental conditions, metrics), system domain (e.g., available hardware, available software, metrics), AI domain (e.g., model architecture, model types, technology, raw data with meta data, metrics); E 1.4 Objectives: Development phase: System domain (metrics, benchmarks, priorities), AI domain (metrics, benchmarks, priorities); operation phase: application domain (metrics, benchmarks, priorities), system domain (metrics, benchmarks, priorities), AI domain (metrics, benchmarks, priorities); maintenance phase: application domain thresholds, system domain thresholds, AI domain thresholds.

\textit{E 2 Combination details (historical view)} The explanation element E2 contains the following explanation elements: E 2.1 Combination description (after training): data set, model, model parameters; application domain data (e.g., image with bounding box); meta-data for data set (e.g., data distribution, data size); post-processing mechanism (e.g., data dimensionality, adjustments, data format changes); quantitative information (e.g., confidence level, ranking position); performance metrics (e.g., ROC curve, PR curve, confusion matrix, robustness); E 2.2 Combination description (during training): Data set, model, model parameters;  meta-data for data set (e.g., data distribution, data size); performance metrics (e.g., ROC curve, PR curve, confusion matrix, loss function, bias/variants function); E 2.3 Combination description (prior to training): raw data; pre-process mechanism (e.g., data cleaning mechanism); feature selection; data set (output); data mining insights (e.g., patterns, statistics).

\textit {E 3 Combination details (simulated production environment)} The explanation element E3 contains the following explanation elements: E 3.1 Combination description (after simulation in production-like environment): data set, model, model parameters; application domain data (e.g., image with bounding box); meta-data for data set (e.g., data distribution, data size); quantitative information (e.g., confidence level, ranking position); performance and cost metrics (e.g., inference time, preprocessing time, precision, robustness); E 3.2 Simulation cases: all simulation cases; applied simulation cases; E 3.3 Description of simulated production environment: virtual application domain; virtual system domain; virtual engine.

\textit{E 4 Combination description (interim and final)} The explanation element E4 contains the following explanation elements: E 4.1 Combination description (combined): data set, model, model parameters; application domain data (e.g., image with bounding box); objective, constraint, metrics achievements; strengths and weaknesses (final achievement); quantitative information (e.g., confidence level, ranking position); differences between the testing view and the evaluation for common metrics.

\textit{E 5 Ranked combinations} The explanation element E5 contains the following explanation elements: E 5.1 Combination summary:  model ID; generation progress; achieved constraints; achieved objective summary; ranking mechanism. 

The explanation elements E 1 through E 5 can be mapped to different domains. The application domain refers to the domain in which the system under design is used. Examples of application domains include manufacturing, healthcare, railroads, building technologies, road traffic, etc. The system domain refers to the domain that provides computational resources. It includes the hardware and software on which the models are executed, such as (industrial) PCs, edge devices, cloud computing, sensor devices (such as cameras), etc. The AI domain is the domain of the artificial intelligence world, including machine learning, data sets, data curation, pipelines, etc.

With the exception of E2, all explanation elements include all three domains (application, system, and AI domain). E2 includes the AI domain only.


\subsubsection{Revised mental model for UT 3 Maintain ML Model}

\begin{figure}[!h]
	\includegraphics[width=1.0\textwidth,angle=0]{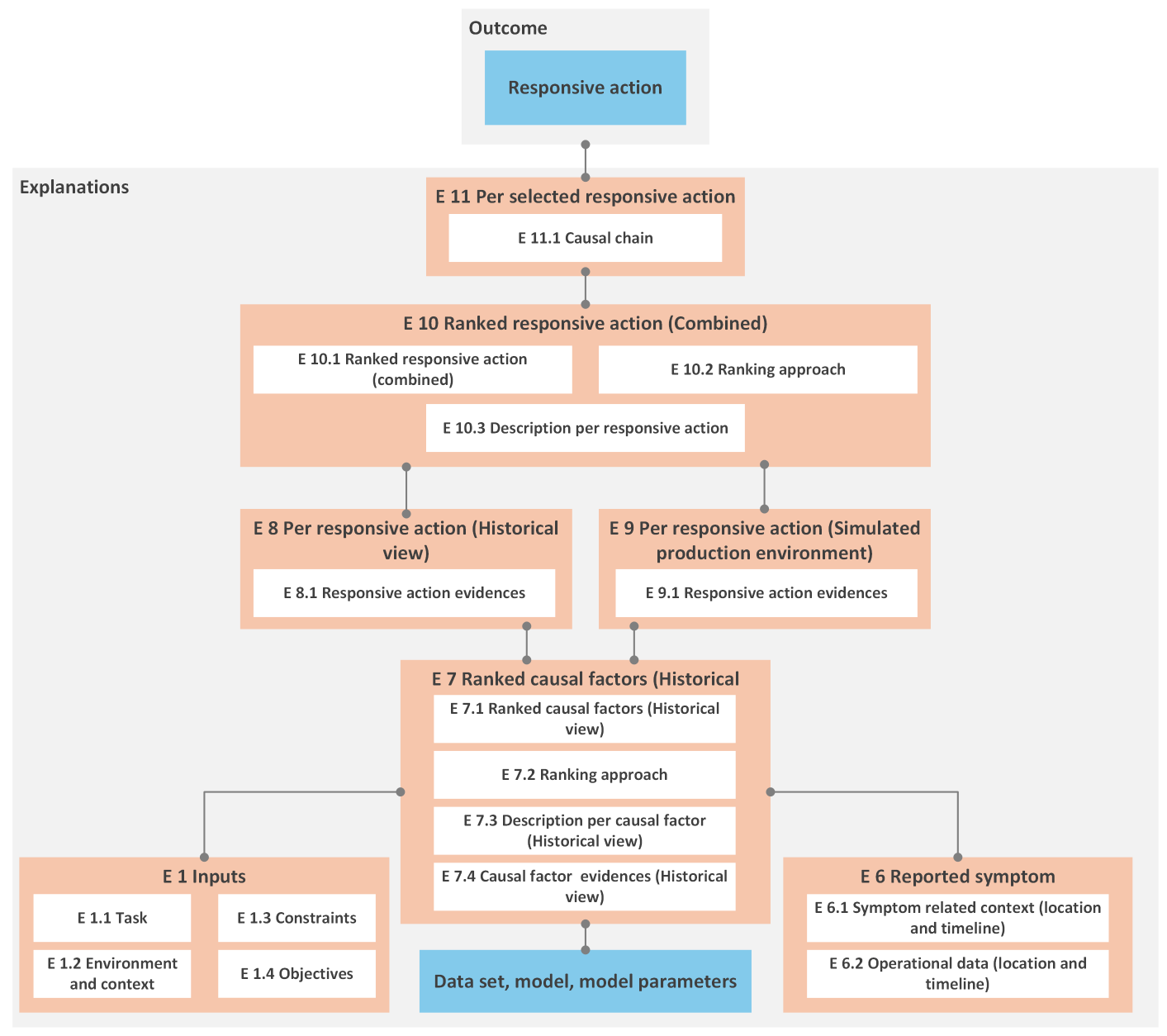}
	\caption{Revised mental model for user task "UT 3 Maintain ML-model"}
	\label{fig:RevisedMentalModel2}
\end{figure}

The revised mental model for explanations for UT 3 (Maintain ML model) contains the following explanation elements (see Figure~\ref{fig:RevisedMentalModel2}).

\textit{E 1 Inputs} The explanation element E 1 is identical with the explanation element E 1 from the UT 2 Generate ML model.

\textit{E 6 Reported symptom} The explanation element E 6 contains the following explanation elements: E 6.1 Symptom related context (location and timeline): Application domain: Changes in application domain (e.g., speed of conveyor belt, lights, change material); system domain: changes in hardware or software (e.g., settings, upgraded OS version, replaced hardware, change of network protocol); AI domain: changes in data (e.g., data drift, out of distribution); E 6.2 Operational data (location and timeline):  Application domain: Deviation of application metrics (e.g., manufacturing throughput), derived deviation patterns of application domain metrics, raw data (historical view) for selected time frame; system domain: deviation system metrics (e.g., end-to-end latency, availability), derived deviation patterns of system domain metrics, deviation patterns across multiple systems (e.g., symptom occurred on IPC 1 and IPC 2 that have the same model deployed); AI domain: deviations of AI metrics (e.g., robustness, accuracy).

\textit{E 7 Ranked causal factors (Historical view)} The explanation element E 7 contains the following explanation elements: E 7.1 Ranked causal factors (historical view): Causal factor name, confidence level; E 7.2 Ranking approach: Mechanism; E 7.3 Causal factor description (historical view): causal factor description; impact on components; projected time; projected ROI; projected outcomes; similar deployments (symptom metrics); cost of not taking an action; E 7.4 Causal factor evidences (historical view): historical cases; quantitative information.

\textit{E 8 Per responsive action (Historical view)} The explanation element E 8 contains the following explanation elements: E 8.1 Responsive action evidences (historical view): historical cases; quantitative information.

\textit{E 9 Per responsive action (Simulated production environment)} The explanation element E 9 contains the following explanation elements: E 9.1 Responsive action evidences (simulated production environment): simulation cases; quantitative information.

\textit{E 10 Ranked responsive actions (Combined)} The explanation element E 10 contains the following explanation elements: E 10.1 Ranked responsive action (combined): Responsive action name; confidence level; E 10.2 Ranking approach: Mechanism; E 10.3 Description per responsive action: Action description; impacted system components; projected time (for repair and down time); projected return-on-investment; projected objectives.

\textit{E 11 Causal chain} The explanation element E 11 contains the following explanation elements: E 11.1 Causal chain: Context; Symptom; causal factor(s); immediate action(s); responsive action(s); projected objectives.

The explanation elements E 1 and E 6 through E 11 can be mapped to different domains. All explanation elements include content from the application domain, the system domain, and the AI domain.


\section{Study 2: Quantitative study}
\label{sec:Study2}


\subsection{Study objective, participant, protocol}
\label{sec:Study2:Goal}

The objective of the study is to understand which explanation content from the revised mental model for explanations is needed for specified explanation intents. Additionally, the study aims to validate the quality of the two mental models for explanations.

We used six screening criteria for the selection of participants (see~Appendix \ref{app:sub:Screener}). We planned to recruit participants from the Siemens organization. The recruitment took place through the personal network of the research team. To reach saturation with a homogeneous study sample, the target sample size is set to twelve participants \cite[p. 7]{Hennink2022-article} \cite[p. 74]{Guest2006-article}.

This study employs a semi-guided interview approach. The study protocol is detailed in the Appendix~\ref{app:sub:protocols}. The HCAI expert conducted the interviews, while two domain experts participated in each session to answer AI-specific questions requiring in-depth domain knowledge.


\subsection{Data analysis}
\label{sec:Study2:Analysis}

We performed the following data analysis for the received answers:

\begin{itemize} 
	\item For step 4: The research team counted how often each explanation content was selected per explanation intent. 
	\item For step 5: The authors calculated the average of all answers. 
	\item For step 6: The answers were grouped by topics. 
	\item For step 7: The research team counted how often each explanation element was selected per explanation intent. 
	\item For step 8: The authors calculated the average of all answers. 
	\item For step 9: The authors calculated the average per intent ranking position. 
	\item For step 10: The research team counted the explanation intents that were excluded. 
	\item For step 11: The research team determined the average rating per explanation intent and ranked the average ratings. 
\end{itemize}


\subsection{Recruited participants}
\label{sec:Study2:RecruitedParticipants}

The research team recruited the same participants from the Siemens organization as in Study 1 (the sample profile is described in subsection~\ref{sec:Study1:RecruitedParticipants}). The interviews took place from January 7 to January 31, 2025. All participants gave explicit consent for the interviews and the use of their anonymized data for publication. The interviews were conducted in English. Each interview lasted from 60 to 90 minutes. All interviews were conducted remotely using Microsoft Teams and were recorded. Participants did not receive any monetary incentive.


\subsection{Study results}
\label{sec:Study2:Results}


\subsubsection{RQ 1: Which explanation content does a data scientist need?}

The participants selected all explanation elements presented in the revised mental model for the user task 'UT 2 Generate ML-Model' (see Table~\ref{tab:Result1.1}).

\begin{table}[!h]
	\footnotesize
	\caption{Selected Explanation Content for Different Explanation Intents in UT 1 Generate ML Model (n = 12). The explanation element group with the highest percentage for each intent is highlighted. (E 1: Input; E 2: Combination Details (Historical View); E 3: Combination Details (Simulated Production Environment); E 4: Combination Description (Interim and Final); E 5: Ranked Combination)}
	\label{tab:Result1.1}
	\begin{tabular}{>{\raggedright\arraybackslash}m{1.8cm} >{\centering\arraybackslash}m{1.6cm} >{\centering\arraybackslash}m{1.6cm} >{\centering\arraybackslash}m{1.6cm} >{\centering\arraybackslash}m{1.6cm} >{\centering\arraybackslash}m{1.6cm} >{\centering\arraybackslash}m{1.6cm}} 
		\hline
		Explanation intent & E 1 & E 2  & E 3 & E 4 & E 5 & Total\\	
		\hline
		Reason 		& 5	&  5	& 3	& 7	& \cellcolor[HTML]{C0C0C0} 11 					&  \cellcolor[HTML]{C0C0C0} 31 \\
		Comparison 	& 4 &  2	& 2	& 7	&  \cellcolor[HTML]{C0C0C0} 9					& 24 \\
		Accuracy 	& 5	&  5	& 2	&  \cellcolor[HTML]{C0C0C0} 6	& 4	& 22\\
		Prediction 	& 3	&  4	&  4	& 4	& \cellcolor[HTML]{C0C0C0} 5 	& 20\\
		Trust 		& 3	&  6	&  6	& \cellcolor[HTML]{C0C0C0} 7	&  \cellcolor[HTML]{C0C0C0} 7 	& 29\\
		\hline
		Total 		& 20	&  22	& 17	& 31	& \cellcolor[HTML]{C0C0C0} 36 & 126\\
		\hline
	\end{tabular}
	Reading example: Eleven (out of twelve) participants selected E 5 as an explanation content that helps data scientists to understand why an outcome was selected (reason intent).
\end{table}

Overall, explanation content E 5 (Ranked combinations) was selected most often (36 times), followed by E 4 (Combination description - Interim and final) (31 times). Participants selected the most explanation content for the explanation intents 'reason' (31) and 'trust' (29). The participants selected all explanation elements presented in the revised mental model for the user task 'UT 3 Maintain ML-Model' (see Table~\ref{tab:Result1.2}).
\begin{table}[!h]
	\footnotesize
	\caption{UT 3 Maintain ML Model - selected explanation content for different explanation intents (n = 12); the explanation element group with the highest percentage per explanation intent is highlighted. E 1 Inputs; E 6 Reported symptom; E 7 Ranked causal factors (Historical view); E 8 Per responsive action (Historical view); E 9 Per responsive action (Simulated production environment); E 10 Ranked responsive actions (Combined); E 11 Causal chain}
	\label{tab:Result1.2}
	\resizebox{\textwidth}{!}{%
		\begin{tabular}{ >{\raggedright\arraybackslash}m{1.8cm} >{\centering\arraybackslash}m{1.4cm} >{\centering\arraybackslash}m{1.4cm} >{\centering\arraybackslash}m{1.4cm} >{\centering\arraybackslash}m{1.4cm} >{\centering\arraybackslash}m{1.4cm} >{\centering\arraybackslash}m{1.4cm} >{\centering\arraybackslash}m{1.4cm} >{\centering\arraybackslash}m{1.4cm}}
			\hline
			Explanation intent & E 1 	& E 6 	& E 7 & E 8 & E 9 & E 10 & E 11 & Total\\
			\hline
			Reason 		& 2		& 2		& 3		& \cellcolor[HTML]{C0C0C0} 8 & 7 & 7 & 1 & 30 \\	
			Comparison 	& 1		& 4		& 4		& \cellcolor[HTML]{C0C0C0} 10 & 9 & 9 & 2 & \cellcolor[HTML]{C0C0C0} 39\\
			Accuracy 	& 1		& 3		& 3		& 3		& 4 	& \cellcolor[HTML]{C0C0C0} 8 	& 0 & 22 \\
			Prediction	& 1		& 3		& 3		& 4		& 5 	& \cellcolor[HTML]{C0C0C0} 11 	& 1 & 28 \\
			Trust 		& 2		& 5		& 6		& \cellcolor[HTML]{C0C0C0} 7 & 5 	&  6 	& 3 &  34\\
			\hline
			Total 		& 7		& 17	& 19	& 32	& 30 & \cellcolor[HTML]{C0C0C0} 41 		& 7 & 153 \\
			\hline
		\end{tabular}%
	}
	\\ Reading example: E 8 was selected eight times as explanation content that helps data scientists to understand why an outcome was selected (reason intent).
\end{table}

Overall, explanation content E 10 was selected most often (41 times), followed by E 8 (32 times) and E 7 (19 times). Participants selected the most explanation content for the explanation intent 'Comparison' (39 times), followed by 'Trust' (34 times). 

Eleven out of twelve participants agree or strongly agree that the presented mental models for explanation help data scientists make confident decisions to select (or not select) a combination (UT 2) or a responsive action (UT 3) (see Table~\ref{tab:Result1.3}).
\begin{table}[!h]
	\footnotesize
	\caption{How well do the explanation contents help the data scientist to make confident decision (n = 12); Statement for mental model for explanations (UT 2): "The explanation content helps the data scientist to make a \underline{confident decision} to select a combination (data model, model, model parameters)"; statement for mental model for explanations (UT 3): "The explanation content helps the data scientist to make a \underline{confident decision} to select a responsive action"}
	\label{tab:Result1.3}
	\begin{tabular}{>{\raggedright\arraybackslash}m{4cm} >{\centering\arraybackslash}m{1.5cm} >{\centering\arraybackslash}m{1.5cm} >{\centering\arraybackslash}m{1.5cm} >{\centering\arraybackslash}m{1.5cm} >{\centering\arraybackslash}m{1.5cm}} 
		\hline
		& 1 					& 2 			& 3				& 4			& 5 \\
		Rated statements								& Strongly disagree 	& Disagree		& Neutral	 	& Agree		& Strongly agree \\
		\hline
		Statement for MM for explanations (UT 2)				& 0 & 0 & 1 & 7 & 4\\
		\hline
		Statement for MM for explanations (UT 3)				& 0 & 0 & 1 & 7 & 4\\
		\hline
	\end{tabular}
	Reading example: Seven (out of twelve) participants rated the statement (see table caption) for mental model for explanations (UT 2) as 'Agree.'
\end{table}

For the first revised mental model (for UT 2 Generate model), participants gave various reasons for their ratings. One participant mentioned that he doesn't rely on the training data and still needs to analyze the dataset for errors (P5). Another participant couldn't find a gap but '[doesn't] go to extremes' (P9), resulting in selecting 'Agree.' Another participant stated that 'in theory, we have performed a systematic search. In practice, there is always a gap' (P12). One participant was unsure whether the intelligent system does the right thing, saying 'It is not clear whether the ranked results and metrics test the right things' (P10). Another participant missed 'absolute values of inputs/objectives' (P1). For another participant, the causality was missing, for instance, between the inputs and the generated models (P2). One participant said, 'I cannot think of another thing that is needed to be added' (P6). Participant P11 noted that the explanation content 'shows multiple dimensions like the data distribution, modal metrics, system metrics.'

For the second revised mental model (for UT 3 Maintain model), participants provided multiple reasons for their ratings. One participant mentioned that the presentation of information is not very user-friendly and causal connections are missing (P2). Another participant noted that the data scientist is not an expert in the application domain, so 'responsive actions should be discussed with the application domain expert' (P4). One participant assumed that the intelligent system might make mistakes, stating, 'During early versions of models, staff needs to be sent to check the cause. Machines can make mistakes' (P5). Another participant made a similar comment: 'There might be root causes that the system does not know' (P7). One participant missed 'a simulation study based on existing data and simulated for a time window that delivers the envisioned impact' (P12). Other participants were more optimistic. One said, 'If I could see these kinds of things, it would help a lot' (P9). Others stated, 'I can't think of anything that is not here' (P6) and 'the information is conclusive' (P1). Participant P11 mentioned that the 'boxes give better understanding, understanding of the cause, of the action, history, and simulated environment where the action is verified. The simulated environment is very powerful.'


\subsubsection{RQ 2 Which explanation content does a data scientist need for selected goals and selected tasks?}

To better answer the questions, we have distinguished five different explanation intents, reflected by five specific questions.

\textit{RQ 2.1 Which explanation helps the data scientist to understand \underline{why} an outcome was selected?}
For UT 2 Generate ML-Model, the explanation content that mostly helps the data scientist understand why an outcome is selected is E 5 'Ranked combinations' (selected by 11 participants), followed by E 4 'Combination description (Interim and final)' (7), E 1 'Input' and E 2 'Combination details (historical view)' (5 each). For UT 3 Maintain ML-Model, the explanation content that mostly helps the data scientist understand why an outcome was selected is E 8 'Per responsive action (Historical view)' (8), equally followed by E 9 'Ranked responsive action (combined)' and E 10 'Ranked responsive actions (combined)' (7 each).


\textit{RQ 2.2 Which explanation content helps the data scientist to understand why an outcome is \underline{better or worse} than another one?}
For UT 2 Generate ML-Model, the explanation content that mostly helps the data scientist understand which outcome is better or worse than another is E 5 'Ranked combinations' (9), followed by E 4 'Combination description (Interim and final)' (7) and E 1 'Input' (4). For UT 3 Maintain ML-Model, the explanation content that mostly helps the data scientist understand which outcome is better or worse than another is E 8 'Per responsive (Historical view)' (10), equally followed by E 9 'Per responsive action (Simulated production environment)' and E 10 'Ranked responsive actions (combined)' (9 each).


\textit{RQ 2.3 Which explanation content do you need to evaluate the \underline{accuracy} of the outcome?}
For UT 2 Generate ML-Model, the explanation content that mostly helps the data scientist evaluate the accuracy of an outcome is E 4 'Ranked combinations' (6), followed by E 1 'Input' and E 2 'Combination details (historical view)' (5 each). For UT 3 Maintain ML-Model, the explanation content that mostly helps the data scientist evaluate the accuracy of an outcome is E 10 'Ranked responsive actions (combined)' (8), followed by E 9 'Per responsive action (Simulated production environment)' (4).


\textit{RQ 2.4 Which explanation content helps the data scientist to \underline{predict} how effective an outcome will be?}
For UT 2 Generate ML-Model, the explanation content that mostly helps the data scientist predict how effective an outcome will be in future use is E 5 'Ranked combination' (5), equally followed by E 2 'Combination details (historical view)', E 3 'Combination details (simulated production environment)', and E 4 'Combination description (Interim and final)' (4 each). For UT 3 Maintain ML-Model, the explanation content that mostly helps the data scientist predict how effective an outcome will be in future use is E 10 'Ranked responsive actions (combined)' (11), followed by E 9 'Per responsive action (Simulated production environment)' (5) and E 8 'Per responsive action (Historical view)' (4).


\textit{RQ 2.5 Which explanation content helps the data scientist to \underline{trust} an outcome?}
For UT 2 Generate ML-Model, the explanation content that mostly helps the data scientist trust an outcome is E 4 'Combination description (Interim and final)' and E 5 'Ranked combinations' (7 each), followed by E 2 'Input' and E 3 'Combination details (simulated production environment)' (6 each). For UT 3 Maintain ML-Model, the explanation content that mostly helps the data scientist trust an outcome is E 8 'Per responsive action (Historical view)' (7), equally followed by E 7 'Causal factor (Historical view)' and E 10 'Ranked responsive actions (combined)' (6 each).


\subsubsection{RQ 3:Which explanation intent has which value for data scientists?}

To better answer the questions, we have split them into two questions.

\textit{RQ 3.1 Which explanation intents have \underline{no value} for data scientists?}
Six participants excluded explanation intents because they did not find them valuable (see Table~\ref{tab:Result7.1}). One participant excluded 'Trust' because 'this is more about understanding than trust' (P3). Another participant made a similar comment when excluding 'Trust': 'It's an objective task' (P7). Participants excluded 'Accuracy' for various reasons: 'Accuracy will be determined later in the life cycle' (P10) and 'Intent is not clear' (P12).

\begin{table}[!h]
	\footnotesize
	\caption{Explanation intents without value for data scientist (n = 12)}
	\label{tab:Result7.1}
	\begin{tabular}{>{\centering\arraybackslash}m{2.3cm} >{\centering\arraybackslash}m{2.3cm} >{\centering\arraybackslash}m{2.3cm} >{\centering\arraybackslash}m{2.3cm} >{\centering\arraybackslash}m{2.3cm}} 
		\hline
		Reason 					& Comparison 			& Accuracy				& Prediction			& Trust \\
		\hline
		0 & 0 & 3 & 1 & 2\\
		\hline
	\end{tabular}
	Reading example: two (out of twelve) participants excluded "Trust" as an explanation intent because it does not have value for them.
\end{table}


\textit{RQ 3.2 Which explanation intents have \underline{the highest value} for data scientists?}
The intent 'Reason' was selected most often as the explanation intent with the highest value for the participants. Participants provided various justifications, such as 'understanding the reason is fundamental' (P3) and 'it sets the foundation to start the journey of improvement and refinement' (P4). P12 justified his selection by stating, 'it is about the causal chain, that is important to me.'

Participants who selected 'Trust' as the intent with the highest value explained it this way: 'For data scientists, it will be important to trust the output of a model, otherwise we don't need it [the model]' and 'trust is essential for any other operation that comes down the road' (P6). Two participants selected 'Prediction' as the intent with the highest value. One participant stated, 'It's tough. I don't care why it works if it works' (P9). One participant selected 'Comparison' as the intent with the highest value, stating, 'Numbers should speak and I make a choice. All other explanations are built on the comparison' (P7).

\begin{table}[!h]
	\footnotesize
	\caption{Rank the following statements; "1" means most important and "5" means least important (n = 12)}
	\label{tab:Result7.2}
	\begin{tabular}{>{\raggedright\arraybackslash}m{6.1cm} | >{\centering\arraybackslash}m{0.9cm} >{\centering\arraybackslash}m{0.9cm} >{\centering\arraybackslash}m{0.9cm} >{\centering\arraybackslash}m{0.9cm} >{\centering\arraybackslash}m{0.9cm}  | >{\centering\arraybackslash}m{0.9cm}} 
		\hline
		Statements expressing an explanation intent		& Rank 1 & Rank 2 & Rank 3 & Rank 4 & Rank 5 & Mean Rank \\
		\hline
		S 1 The explanation content should help the data scientist to understand \underline{why} an outcome was identified. ("Reason" intent)							& 4 & 4 & 2 & 2 & 0 & \cellcolor[HTML]{C0C0C0} 2.2\\
		\hline
		S 2 The explanation content should help the data scientist to understand why a certain outcome is \underline{better or worse} than another one. ("Comparison" intent)& 2 & 1 & 6 & 2 & 1 & 2.9\\
		\hline
		S 3 The explanation content should help the data scientist to evaluate the \underline{accuracy} of the identified outcome. ("Accuracy" intent)	& 1 & 2 & 2 & 2 & 2 & 3.2\\
		\hline
		S 4 The explanation content should help the data scientist to \underline{predict how effective} the identified outcome will be. ("Predictability" intent) & 2 & 3 & 2 & 3 & 1 & 2.8\\
		\hline
		S 5 The explanation content should help the data scientist to \underline{trust} the identified outcome.("Trust" intent)									& 3 & 2 & 0 & 3 & 2 &  2.9\\		
		\hline
	\end{tabular}
	Reading example: The participants selected statement S 1 four times for rank 1, four times for rank 2, twice for rank 3, twice for rank 4, and none for rank 5. The mean rank is 2.2. 
\end{table}



\section{Discussion, limitations, and future work}
\label{sec:Findings}


\subsection{Discussions}

We conducted a mixed-method study, incorporating both qualitative and quantitative approaches, to understand which explanation content data scientists need to make confident decisions regarding the outcomes of an intelligent system designed to generate, deploy, and maintain ML models. For the user tasks UT 2 (Generate ML model) and UT 3 (Maintain ML model), the research team, comprising an HCI researcher and two data scientists, created initial mental models for explanations per user task.

To perform the study, we deployed a revised method, using a mental model for explanations, that includes the following steps (see also Figure~\ref{fig:ResearchApproach}): 1) Create a raw mental model for explanations (HCAI researcher and one data scientist). 2) Validate the raw mental model for explanations (with the HCAI researcher and both data scientists, AKA the 'research team'), and improve it (qualitative study). The improved mental model is called the 'initial mental model.' 3) Validate the initial mental model with representatives of the target user group (n = 12; qualitative study). 4) Revise the initial mental model (research team), called the 'revised mental model.' 5) Create a UI mockup, based on the revised mental model; 6) Validate the mental model and understand which explanation content is needed for which explanation intents with representatives of the target user group (n = 12; quantitative study). In previous studies \cite{Degen2023-inproceedings,Degen2024-inproceedings}, steps 2, 3, and 4 were not performed. The steps were introduced to increase the quality of the mental model for explanations. 

The goal of the first, qualitative study (step 3) was to validate the initial mental models for explanations and identify areas for improvement. We evaluated the initial models by asking the following questions: Q1) Which explanation content is missing, and why? Q2) Which explanation content is not needed, and why? Q3) Which explanation content should be restructured or reorganized, and why? After collecting the feedback, we discussed it in a workshop using open coding and agreed as a team on the necessary modifications to the initial mental models. The goal of the second, quantitative study (step 6) was to understand which explanation content helps data scientists make confident decisions to select an outcome. The explanation intents under research are: reason (ability to understand why an outcome was identified), comparison (understanding which outcome is better or worse than another outcome), accuracy (validating the accuracy of an outcome), predictability (predicting how effective an outcome will be in future use), and trust (trusting an outcome). We also asked participants to identify explanation intents that do not have value for them and to rank the explanation intents that do. We derived the following findings from the studies:

\textit{Finding 1 (Explanation domains): Explanation content for data scientists comes from the application domain, system domain, and AI domain.} 
With one exception (E 2 'Combination details (historical view)'), all explanation elements contain content from the application domain, the system domain, and the AI domain. The inclusion of explanation content from different domains is not new \cite{Degen2023-inproceedings,Degen2024-inproceedings}. This study confirms that the results of previous studies are consistent. It is somewhat surprising that data scientists with a focus on AI also need explanations from the application and system domains.

\textit{Finding 2 (Explanation structures): Explanation content can be complex and should be organized sequentially and/or in hierarchies.}
As demonstrated in this study, explanations can be complex. To make complex explanation content accessible and digestible for users, it can be organized in a sequence that represents a causal relationship (also known as 'show your work' explanations \cite{Degen2023-inproceedings}). The use of causal explanations is not new \cite{Carloni2023-XAI-Rolecausality-arxiv-article,Chou2023-XAI-Counterfactual-Causal-InfoFusion-article,Beckers2022-CausalExplanations-inproceedings}. Alternatively, complex explanation content can be organized hierarchically, allowing data scientists to access details step-by-step as needed. This introduces a new explanation property, \textit{explanation structure}, which can be sequential and/or hierarchical.

\textit{Finding 3 (Explanation content types): Explanation content includes context, inputs, evidence, attributes, ranked list, interim results, efficacy principle, and input/output relationships.}
Data scientists expressed the need to review content related to model generation and model maintenance. The following types of content have been requested: environmental conditions and application domain (context), raw data (inputs), input parameters (inputs), the actual data set (evidence), symptoms and root causes (causal factors), metadata of data sets (attributes), the ranked list of models or responsive actions (ranked list), interim models or root causes based on historic data (interim results), the causal connection between symptom and root cause as well as root cause and responsive action (efficacy principle), and the input/output relationship that allows changes to the inputs to achieve certain outcomes (input/outcome relationship). It is important to note that these content types should be organized as a causal story (see Finding 4). These new types of explanation content can be added to the list of explanation properties as \textit{explanation content types}.

\textit{Finding 4 (Explanation structure): Explanation content should be organized as a causal story.}
This study confirms that explanation content should be organized in a 'causal structure' to tell a causal story. Data scientists prefer to see the higher-level picture first (causal story) and then delve deeper into the details. It is important to emphasize that the 'causal structure' of explanations is a \textit{causality assertion}. It is possible that some explanation elements in the 'causal story' do not have a causal relationship with adjacent elements. Data scientists need to validate that the elements have a causal relationship, which supports the validity of the outcome. The finding that explanation content needs to be organized as a causal story is not new and aligns with findings from other studies \cite{DegenBudnik2021-inproceedings,Degen2023-inproceedings,Degen2024-inproceedings}.

\textit{Finding 5 (Explanation questions): Standardized explanation questions ensure complete coverage of explanation needs.}
In contrast to our previous studies \cite{Degen2023-inproceedings,Degen2024-inproceedings} using mental models for explanations, we used predefined explanation questions (see Section~\ref{sec:ApplicationDomain:UserRole}) in this study. These predefined questions relate to question banks with standardized questions \cite{Liao2020-QuestionBank-CHI-inproceedings,Schmude2025-QuestionBank-IJHCS-article}. Standardized explanation questions expedite the research because the explanation questions do not need to be identified prior to the task analysis. They also remove an unwanted element of subjectivity. Furthermore, the use of standardized questions ensures that we completely cover the explanation needs for each user role and user task. Using standardized explanation questions as part of the task analysis to identify user role-specific and user task-specific explanation content is a novel approach in the context of XAI.

\textit{Finding 6 (Explanation methodology): Refining mental models for explanations increases significantly its quality.}
In contrast to our previous studies \cite{Degen2023-inproceedings,Degen2024-inproceedings} using no validation of the initial mental model, we introduced a validation and revision step that led to a revised mental model for explanations. By comparing the content differences between the initial (see Section~\ref{sec:Study1:Preparation}) and the revised mental model (see Section~\ref{sec:Study1:ResultsMM}), we found that the validation and revision steps significantly increased the quality of the mental model for explanations and hence increased the trust in understanding which explanation content is needed for which explanation intent. The entire approach to create and refine mental models for explanations is a novel approach in the context of XAI.


\subsection{Limitations and future work}

This research has some limitations. One limitation is the reuse of the same participants for both studies. Due to research constraints and economic considerations, we chose to involve the same participants for both studies. This likely introduced bias in the participants' views during the second, quantitative study.

Another limitation is the review and assessment of the mental models for explanation based on interviews. A behavioral study, such as a formative or summative usability test, would provide additional insights into the need for (context-sensitive) explanation content and evaluate whether the identified explanation content is indeed effective, efficient, and satisfactory in accordance with ISO 9241-110 \cite{ISO9241-110-2020} for our target user group.

Mental models allow us to consider a broad spectrum of effective explanations and help avoid the trap of thinking too narrowly or technology-focused when it comes to explanations. They assist in designing AI systems that 'afford effective explanations' \cite[p. 75]{Carroll2022-article} before cost-intensive research and development begins. When considering different user groups for the same AI system, their mental models are likely different yet interconnected. This is known as a shared mental model \cite{Andrews2023-article}. Research on shared mental models for explainability could derive connected explanations that simplify communication between different user roles. Such research would help identify user role-specific explanations and shared explanations. Another topic for future work is the presentation of a confidence level.


\section{Acknowledgment}
\label{Acknowledgment}
The authors thank all participants for sharing their time and insights with us.

\section{Conflict of interest}
The authors declare that they have no competing financial interests or personal relationships that could have appeared to influence the work reported in this paper.

\section{Funding}
This research was funded by Siemens Corporation. 


%
%
%
\bibliographystyle{splncs04}
\bibliography{XAI_DS_MAthilda_v4.bib}


\section{Appendix}
\label{app:Appendix}


\subsection{Initial mental models}
\label{app:sub:InitialMM}

\begin{figure}
	\includegraphics[width=1.0\textwidth,angle=0]{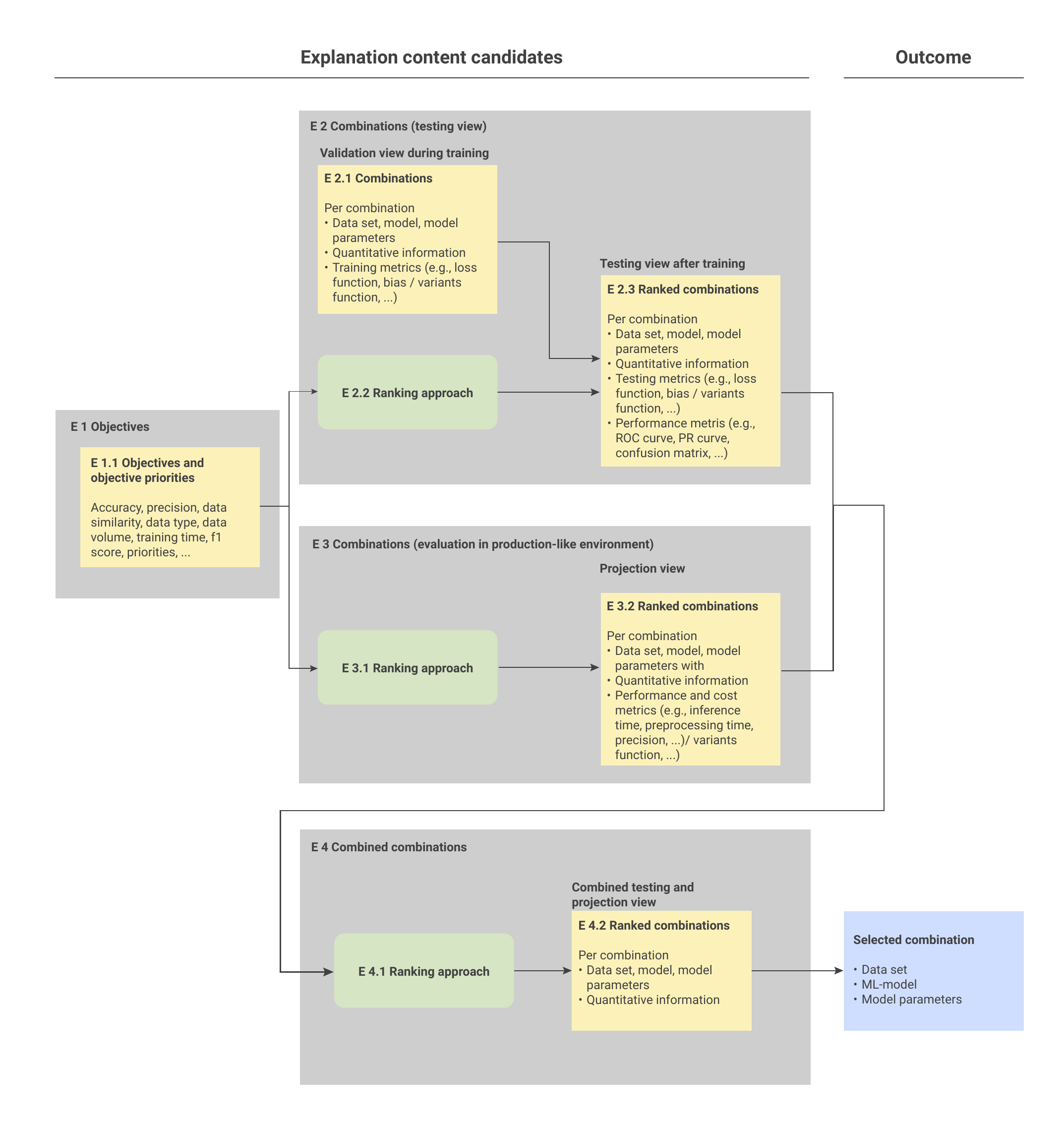}
	\caption{Initial mental model for user task "UT 2 Generate ML-model"}
	\label{fig:Task1_InitialGenerateMLModel}
\end{figure}

\begin{figure}
	\includegraphics[width=1.0\textwidth,angle=0]{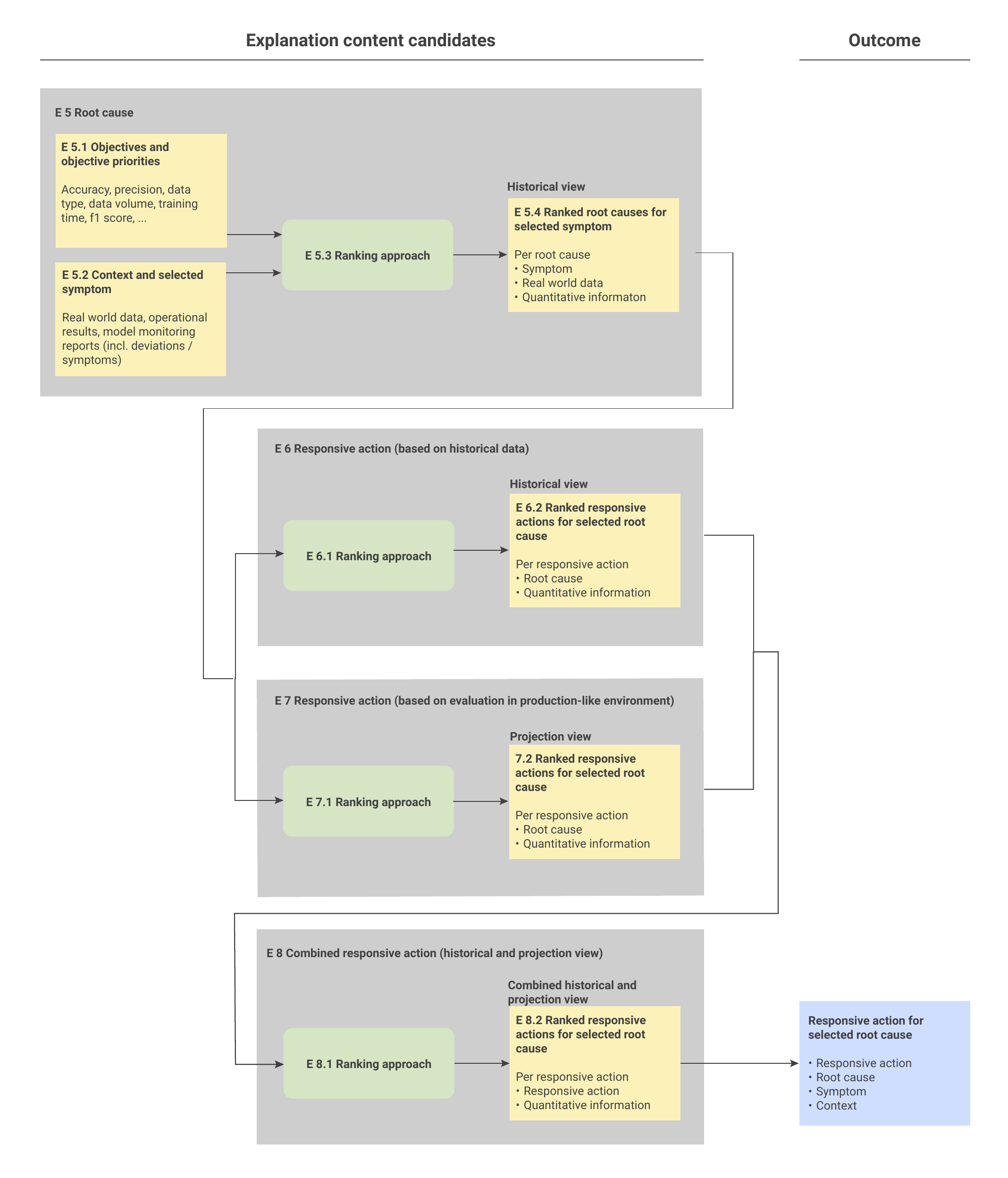}
	\caption{Initial mental model for user task "UT 3 Maintain ML-model"}
	\label{fig:Task2_InitialMonitorMLModel}
\end{figure}


\subsection{Screener criteria}
\label{app:sub:Screener}

We used the following screener questions:

\begin{itemize}
	\item C1: How many years of industrial experience do you have training, testing, and deploying ML frameworks? (years) 
	\item C2: Have you been involved in data selection? (yes/no)
	\item C3: Have you been involved in model training? (yes/no)
	\item C4: Have you been involved in model testing? (yes/no)
	\item C5: Have you been involved model fine tuning? (yes/no)
	\item C6: Have you been involved in model integration / deployment? (yes/no)
	\item C7: Have you been involved in model maintenance? (yes/no)
\end{itemize}

To qualify for the interview, participants needed to meet two criteria: they had to answer question C1 with '3 years' or more and provide at least four 'yes' answers to questions C2 through C7.


\subsection{User task specific explanation needs}
\label{app:sub:UserTask}

\paragraph{User task 1: DS (re) defines objectives (incl. scope, technical objectives, metrics, requirements, ranking, project constraints)}

\begin{itemize}
	\item User's intent: Define criteria for filtering and ranking combinations
	\item Input: Objective value per objective category
	\item Output: Defined objectives
	\item Explanation needs: none
\end{itemize}


\paragraph{User task UT 2: DS (re) selects combination}

\begin{itemize}
	\item Intent: Select the combination with the highest combined criteria value
	\item Input: Defined objectives
	\item Output: Selected combination (one out of multiple suggested combination)
	\item Explanation needs
	\begin{itemize}
		\item EQ1: Which explanation content do you need to understand \underline{how} the outcome was determined?
		\begin{itemize}
			\item Step 1: Identify all combinations that meet the objectives.
			\item Step 2: Assess the combined criteria for each combination.
			\item Step 3: Sort the combinations by their Combined Criteria value.
			\item Step 4: Choose the combination with the highest combined criteria value.
		\end{itemize}
		\item EQ2: Which explanation content do you need to understand \underline{why} the outcome was determined?
		\begin{itemize}
			\item Combination with highest Combined Criteria Value; level of compliance with defined objectives.
		\end{itemize}
		\item EQ3: Which explanation content do you need to understand the \underline{accuracy} of the outcome?
		\begin{itemize}
			\item Looked at Combined Criteria Value.
		\end{itemize}
		\item EQ4: Which explanation content do you need to \underline{predict} the effectiveness of the outcome?
		\begin{itemize}
			\item Looked at Future Data Similarity (fd) value.
		\end{itemize}
	\end{itemize}
\end{itemize}


\paragraph{User task UT 3: Maintain ML-model}

\begin{itemize}
	\item Intent: Keep the model's accuracy above / below a defined threshold
	\item Inputs
	\begin{itemize}
		\item Defined objectives (result of UT 1)
		\item Real world data (input of P-ST 2)
		\item Operational results (result of P-ST-2)
		\item Model monitoring reports containing deviations against defined objectives (result of P-ST-2)
	\end{itemize}
	\item Output: Ranked responsive actions (incl. modify selected combination, replace selected combination)
	\item Explanation needs
	\begin{itemize}
		\item EQ1: Which explanation content do you need to understand \underline{how} the outcome was determined?
		\begin{itemize}
			\item Step 1: Identify symptom (identified deviation); rank symptoms; select highest ranked symptom
			\item Step 2: For selected symptom: Identify root causes; rank root causes; select highest ranked root cause
			\item Step 3: For selected root cause: Identify responsive actions; rank responsive actions; select highest ranked responsive actions
		\end{itemize}
		\item EQ2: Which explanation content do you need to understand \underline{why} the outcome was determined?
		\begin{itemize}
			\item The outcome was determined based on the highest ranked responsive actions.
		\end{itemize}
		\item EQ3: Which explanation content do you need to understand the \underline{accuracy} of the outcome?
		\begin{itemize}
			\item Examining the effectiveness of the selected responsive actions.
		\end{itemize}
		\item EQ4: Which explanation content do you need to \underline{predict} the effectiveness of the outcome?
		\begin{itemize}
			\item Looked at Future Data Similarity (fd) value and historical success of the selected responsive actions in similar situations. Additionally, the relevance of the identified root cause to the symptom and the appropriateness of the responsive action to the root cause could also be considered.
		\end{itemize}
	\end{itemize}
\end{itemize}


\subsection{Study protocols}
\label{app:sub:protocols}

\subsubsection{Protocol of study 1 (qualitative)}

\begin{itemize}
	\item Step 1: Research scope
	\item Step 2: Job experience of participant
	\item Step 3: Introduction into "Intelligent system to generate, deploy, and maintain ML-models" 
	\item Step 4: Introduction into the mental model of explanation for user task 1	"Generate ML-model"
	\item Step 5: For the prepared mental model representation for user task 1: Which explanation content is missing and why?
	\item Step 6: For the prepared mental model representation for user task 1: Which explanation content is not necessary and why?
	\item Step 7: For the prepared mental model representation for user task 1: How should the mental model representation be rearranged and why?
	\item Step 8: For the prepared mental model representation for user task 1: Please rate the following statement: "The explanation content candidates help the data scientist to make a confident decision selecting an effective combination." (answer options: 5-point Likert scale)
	\item Step 9: Introduction into the mental model of explanation for user task 2 "Maintain ML-model"	
	\item Step 10: For the prepared mental model representation for user task 2: Which explanation content is missing and why?
	\item Step 11: For the prepared mental model representation for user task 2: Which explanation content is not necessary and why?
	\item Step 12: For the prepared mental model representation for user task 2: How should the mental model representation be rearranged and why?
	\item Step 13: For the prepared mental model representation for user task 2: Please rate the following statement: "The explanation content candidates help the data scientist to make a confident decision selecting an effective responsive action." (answer options: 5-point Likert scale)
\end{itemize}


\subsubsection{Protocol of study 2 (quantitative)}

\begin{itemize}
	\item Step 1: Research scope
	\item Step 2: Introduction into "Intelligent system to generate, deploy, and maintain ML-models" 
	\item Step 3: Introduction into the revised mental model for explanation for user task 1 "Generate ML-model"; the introduction contained the revised mental model and mockup screen shots including entering models inputs (see Appendix~\nameref{app:Mockup-inputs}), generating models (see Appendix~\nameref{app:Mockup-generate}), and selecting a model (see Appendix~\nameref{app:Mockup-selectmodel}).
	\item Step 4: Questions about the explanation content for different explanation intents for user task 1 (in randomized order per study participant)
	\begin{itemize}
		\item Which explanation helps the data scientist to understand \underline{why} an outcome was selected?
		\item Which explanation content do you need to evaluate the \underline{accuracy} of the outcome?
		\item Which explanation content helps the data scientist to predict how \underline{effective} of an outcome will be?
		\item Which explanation content helps the data scientist to understand why an outcome is \underline{better or worse} than another one?
		\item Which explanation content helps the data scientist to \underline{trust} an outcome?
	\end{itemize}
	\item Step 5: Please rate the following statement: "The explanation content help the data scientist to make a \underline{confident decision} to select (or not to select) a combination." (answer options: 5-point Likert scale)
	\item Step 6: Please provide a reason for the selected rating.
	\item Step 7: Introduction into the revised mental model of explanation for user task 3 "Maintain ML-model"; the introduction contained the revised mental model and mockup screen shots including selecting a responsive action (see Appendix~\nameref{app:Mockup-selectaction}).
	\item Step 8: Questions about the explanation content for different explanation intents for user task 3 (in randomized order per study participant)
	\begin{itemize}
		\item Which explanation helps the data scientist to understand \underline{why} an outcome was selected?
		\item Which explanation content do you need to evaluate the \underline{accuracy} of the outcome?
		\item Which explanation content helps the data scientist to predict how \underline{effective} of an outcome will be?
		\item Which explanation content helps the data scientist to understand why an outcome is \underline{better or worse} than another one?
		\item Which explanation content helps the data scientist to \underline{trust} an outcome?
	\end{itemize}
	\item Step 9: Please rate the following statement: "The explanation content helps the data scientist to make a \underline{confident decision} to select (or not to select) a responsive action." (answer options: 5-point Likert scale)
	\item Step 10: Please provide a reason for the selected rating.
	\item Step 11: Please select the explanation intents that do not have value for you.
	\item Step 12: Please select the explanation intents that are relevant for you and rank the selected once (1: highest priority).
\end{itemize}

\newpage


\subsection{Mockup}


\subsubsection{Enter model inputs}
\label{app:Mockup-inputs}

\begin{figure}[h!]
	\includegraphics[width=1.5\textwidth,angle=90]{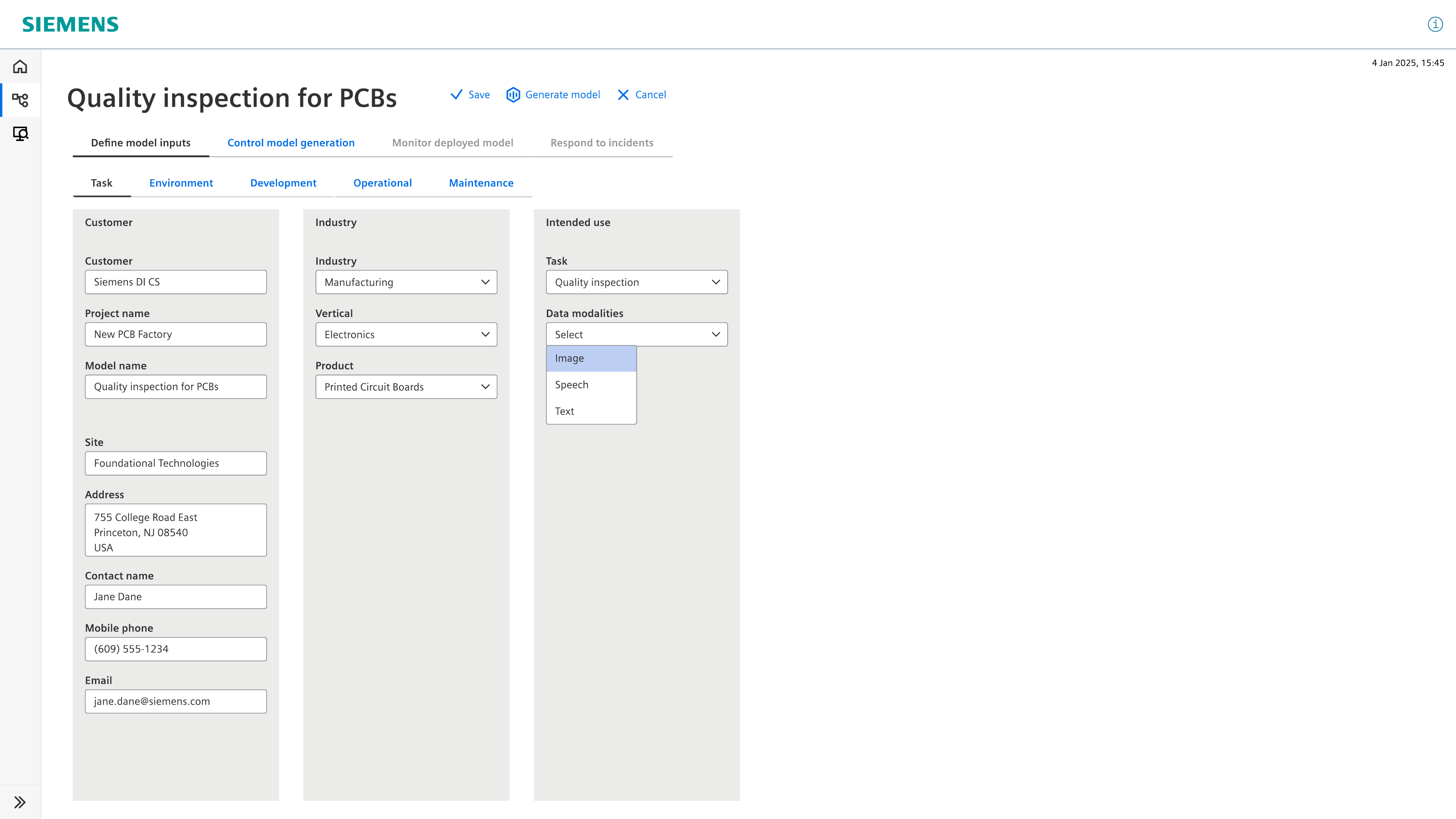}
	\caption{For user task "UT 2 Generate ML-model"; enter model inputs 1}
	\label{fig:Figma_Input1}
\end{figure}

\begin{figure}[h!]
	\includegraphics[width=1.5\textwidth,angle=90]{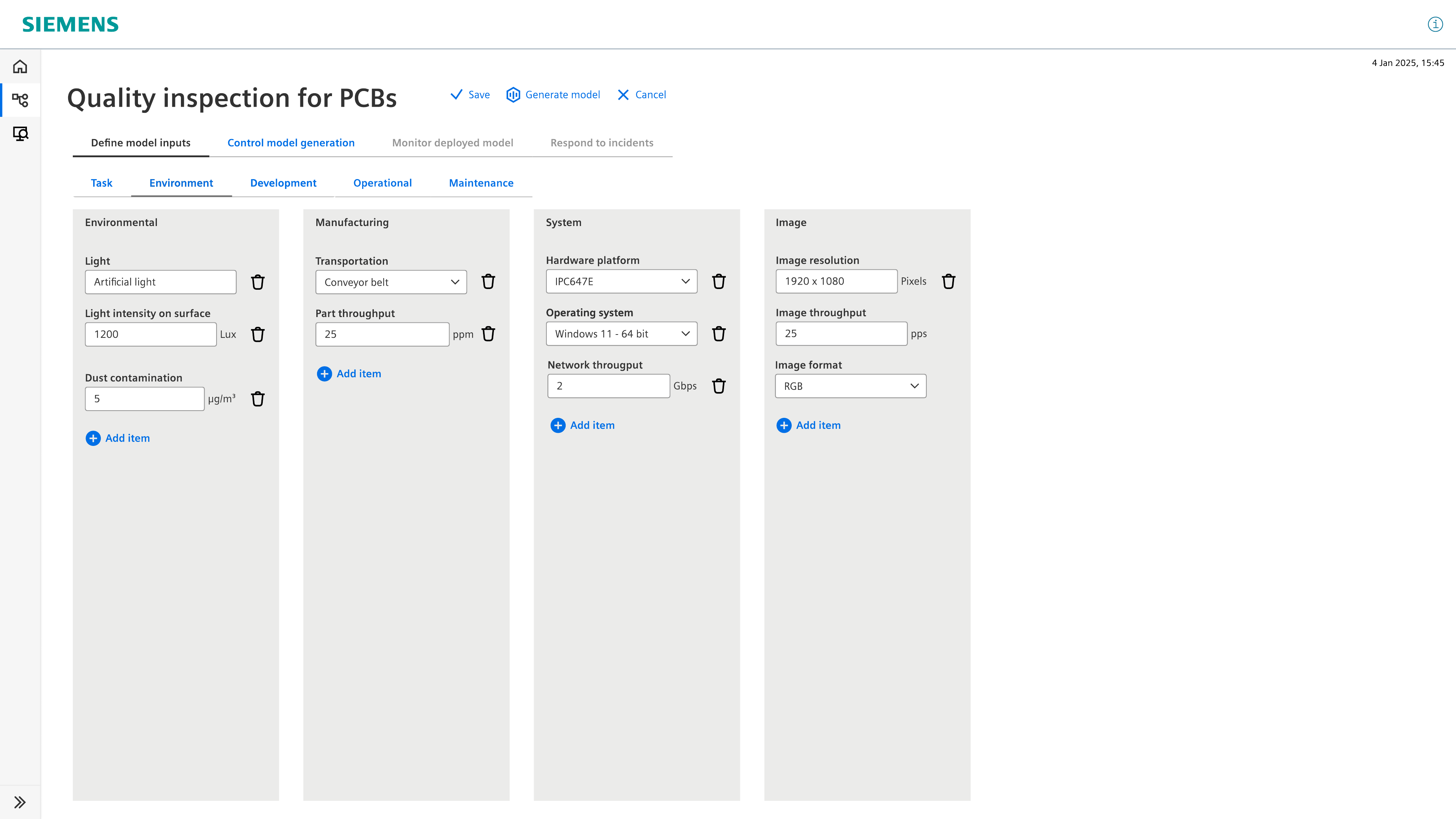}
	\caption{For user task "UT 2 Generate ML-model"; enter model inputs 2}
	\label{fig:Figma_Input2}
\end{figure}

\begin{figure}[h!]
	\includegraphics[width=1.5\textwidth,angle=90]{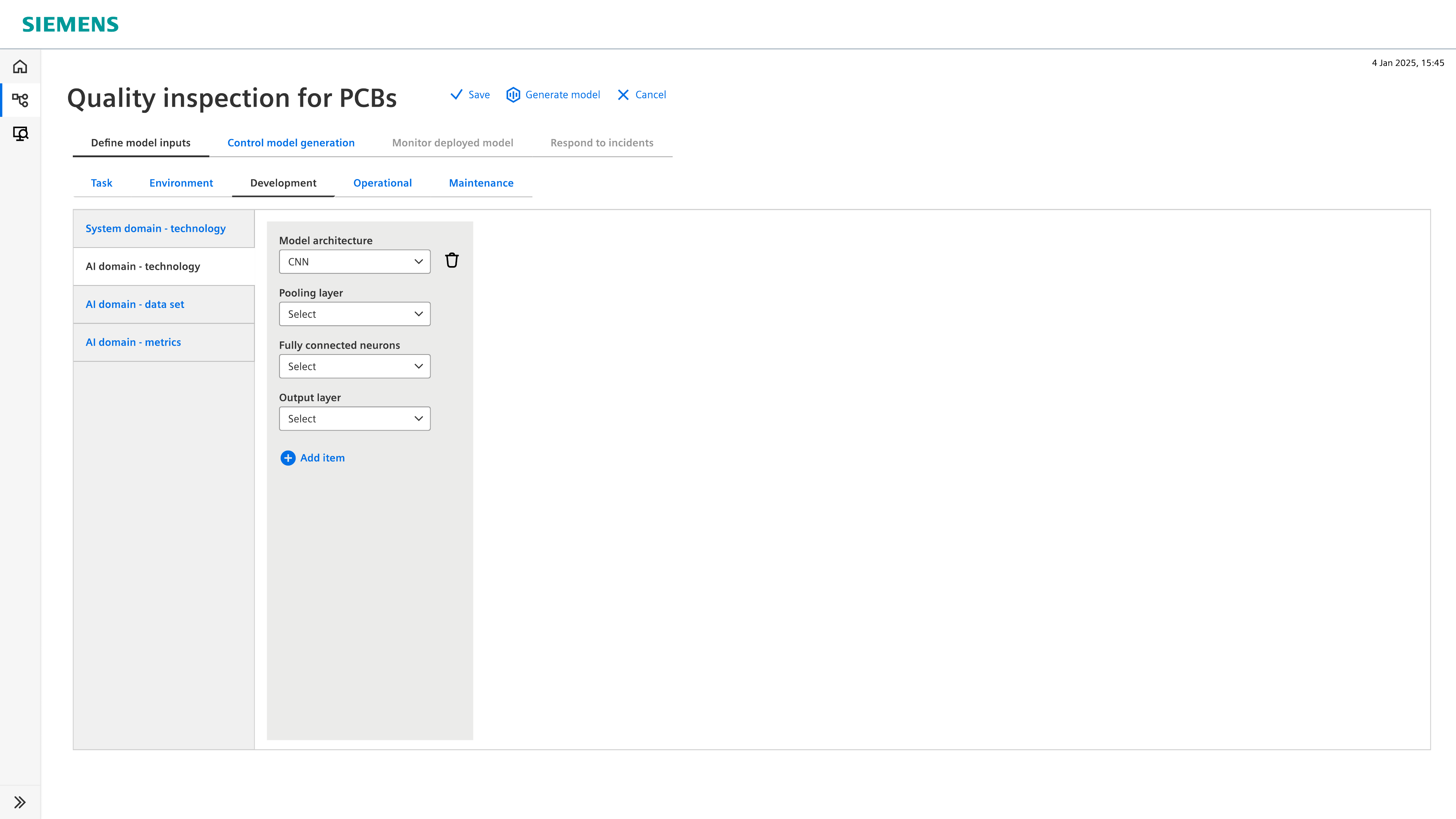}
	\caption{For user task "UT 2 Generate ML-model"; enter model inputs 3}
	\label{fig:Figma_Input3}
\end{figure}

\begin{figure}[h!]
	\includegraphics[width=1.5\textwidth,angle=90]{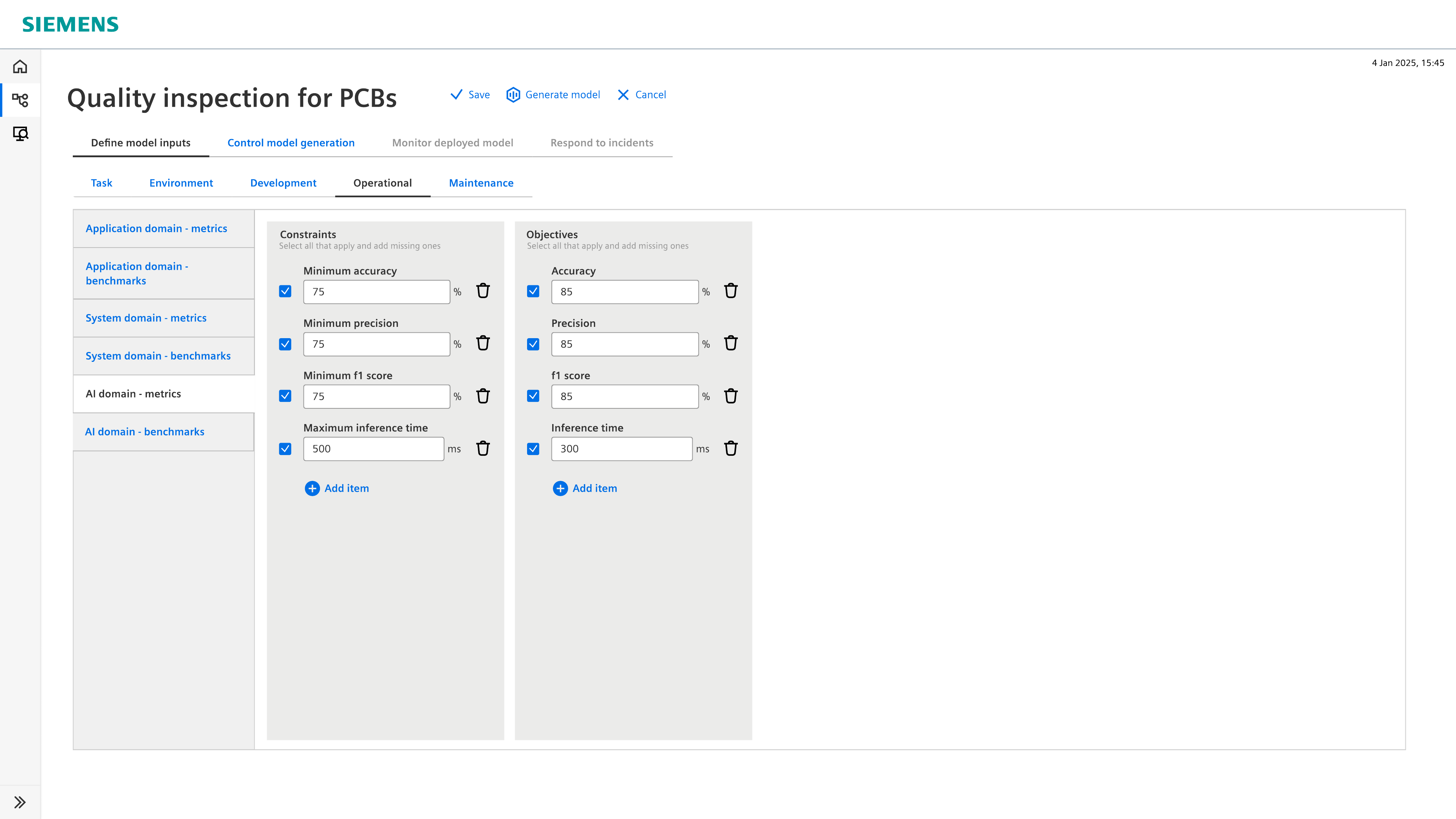}
	\caption{For user task "UT 2 Generate ML-model"; enter model inputs 4}
	\label{fig:Figma_Input4}
\end{figure}

\begin{figure}[h!]
	\includegraphics[width=1.4\textwidth,angle=90]{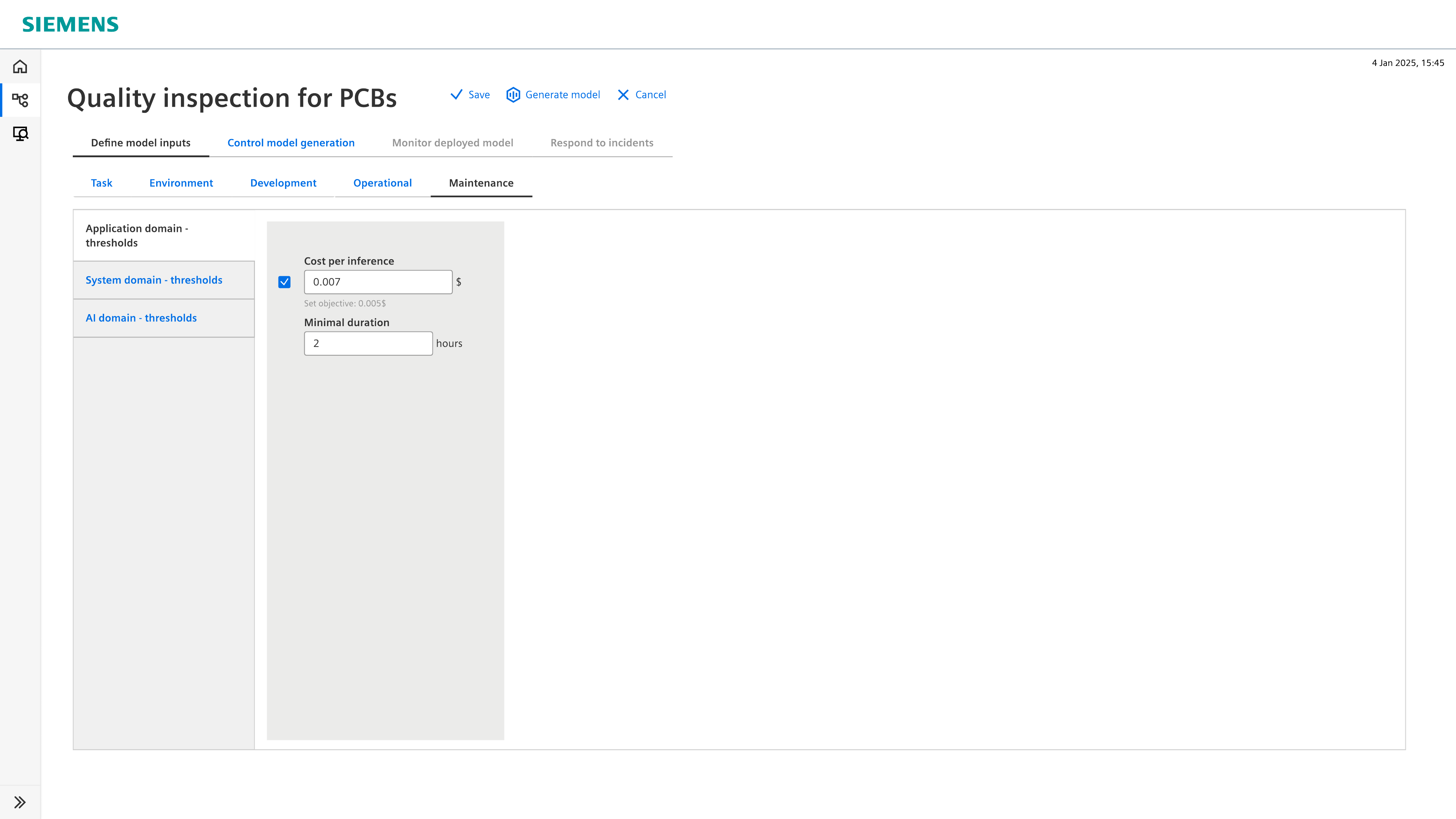}
	\caption{For user task "UT 2 Generate ML-model"; enter model inputs 5}
	\label{fig:Figma_Input5}
\end{figure}



\subsubsection{Generate models}
\label{app:Mockup-generate}

\begin{figure}[h!]
	\includegraphics[width=1.5\textwidth,angle=90]{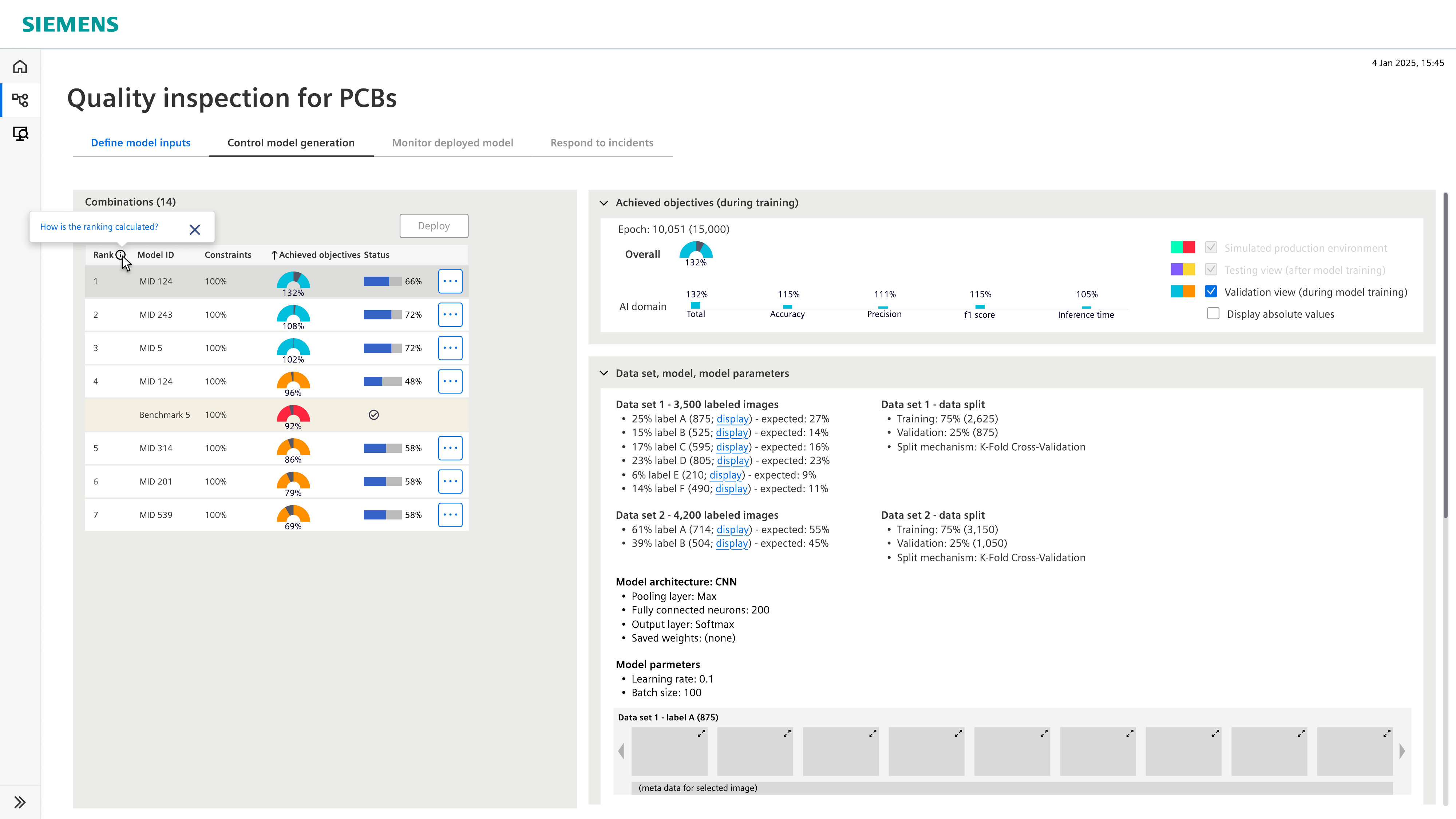}
	\caption{For user task "UT 2 Generate ML-model"; monitoring model generation 1}
	\label{fig:Figma_Generation}
\end{figure}



\subsubsection{Select generated model}
\label{app:Mockup-selectmodel}

\begin{figure}[h!]
	\includegraphics[width=1.5\textwidth,angle=90]{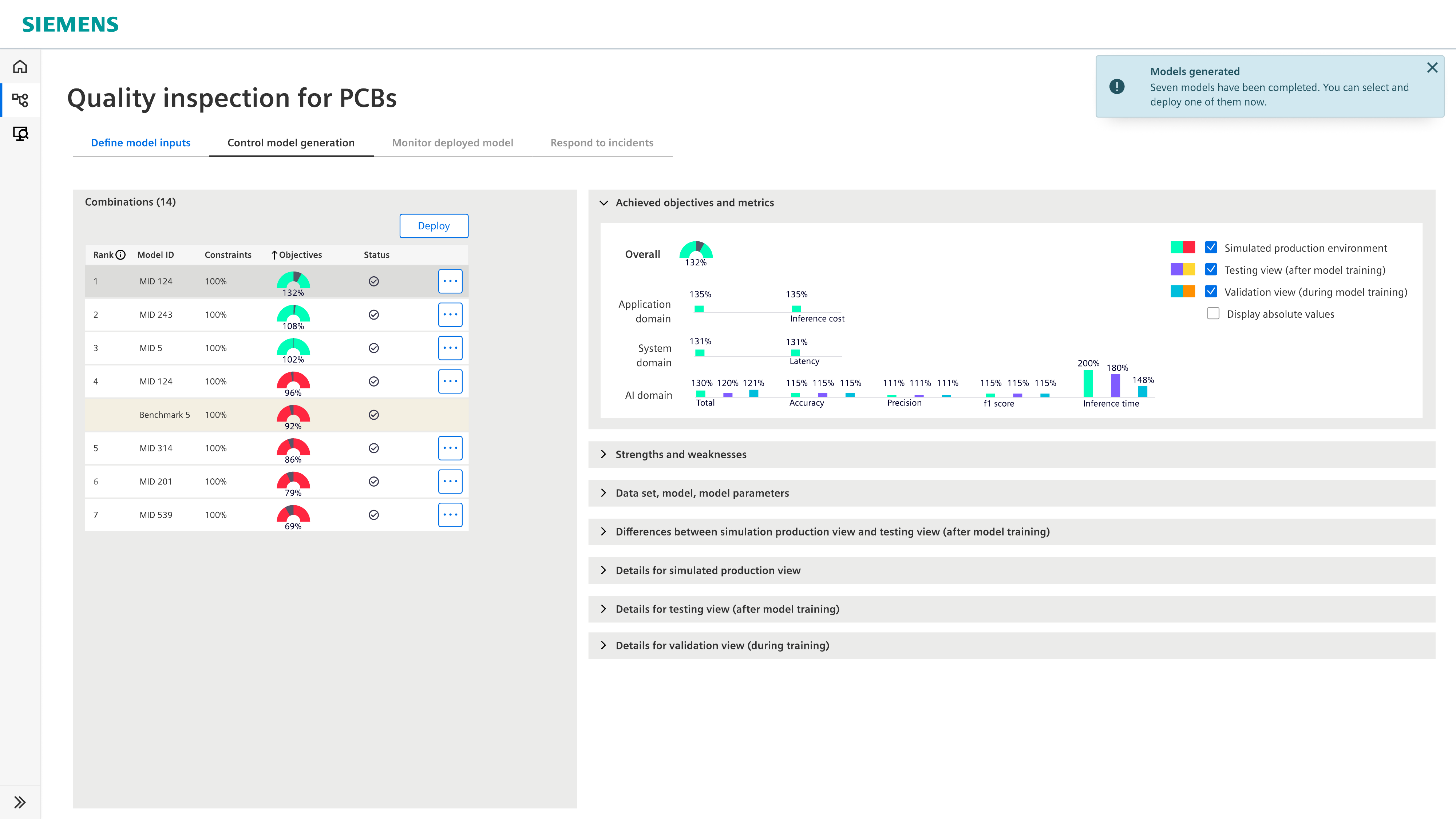}
	\caption{For user task "UT 2 Generate ML-model"; select model 1}
	\label{fig:Figma_Select1}
\end{figure}

\begin{figure}[h!]
	\includegraphics[width=1.5\textwidth,angle=90]{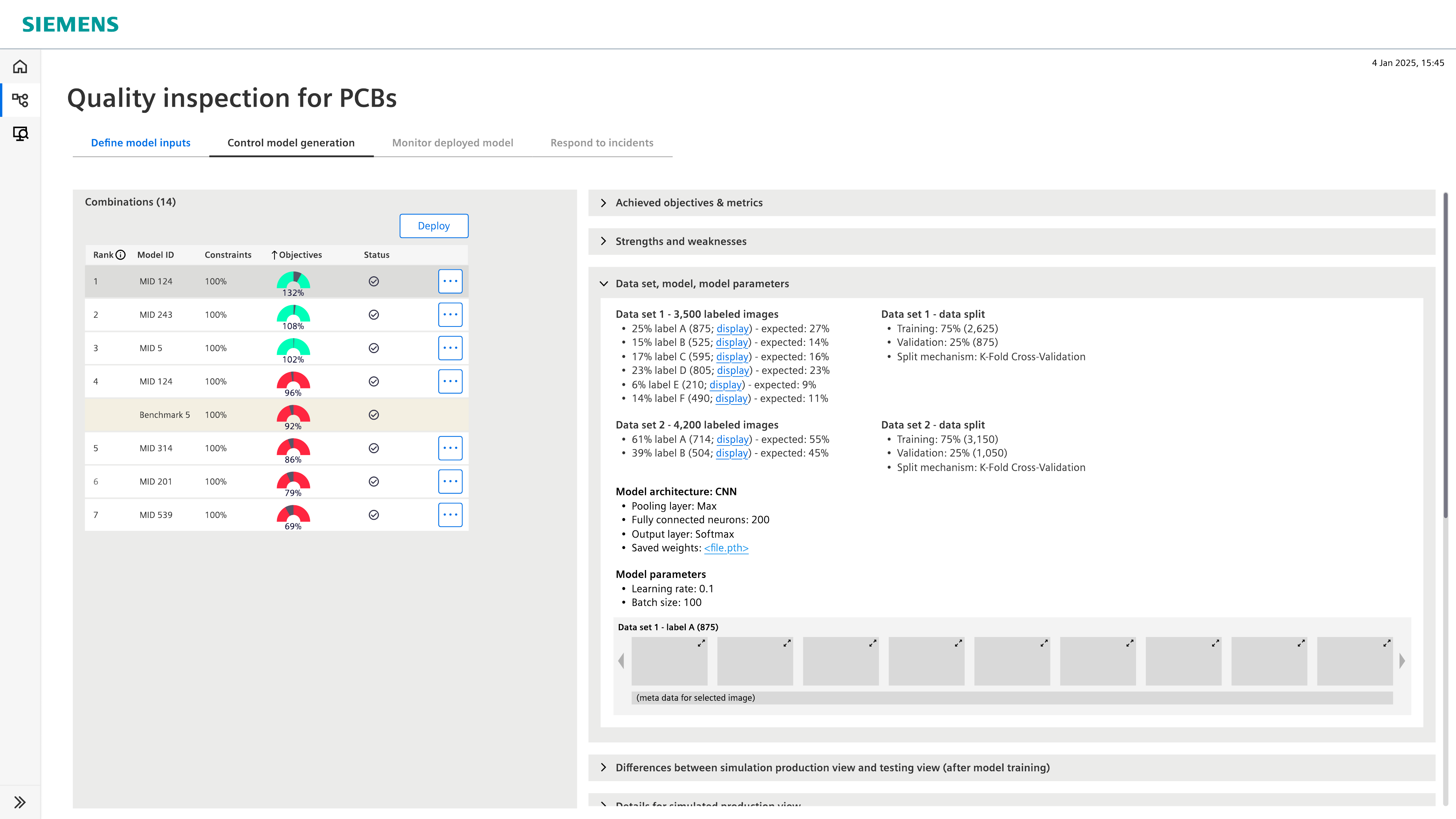}
	\caption{For user task "UT 2 Generate ML-model"; select model 2}
	\label{fig:Figma_Select2}
\end{figure}



\subsubsection{Select responsive action}
\label{app:Mockup-selectaction}

\begin{figure}[h!]
	\includegraphics[width=1.5\textwidth,angle=90]{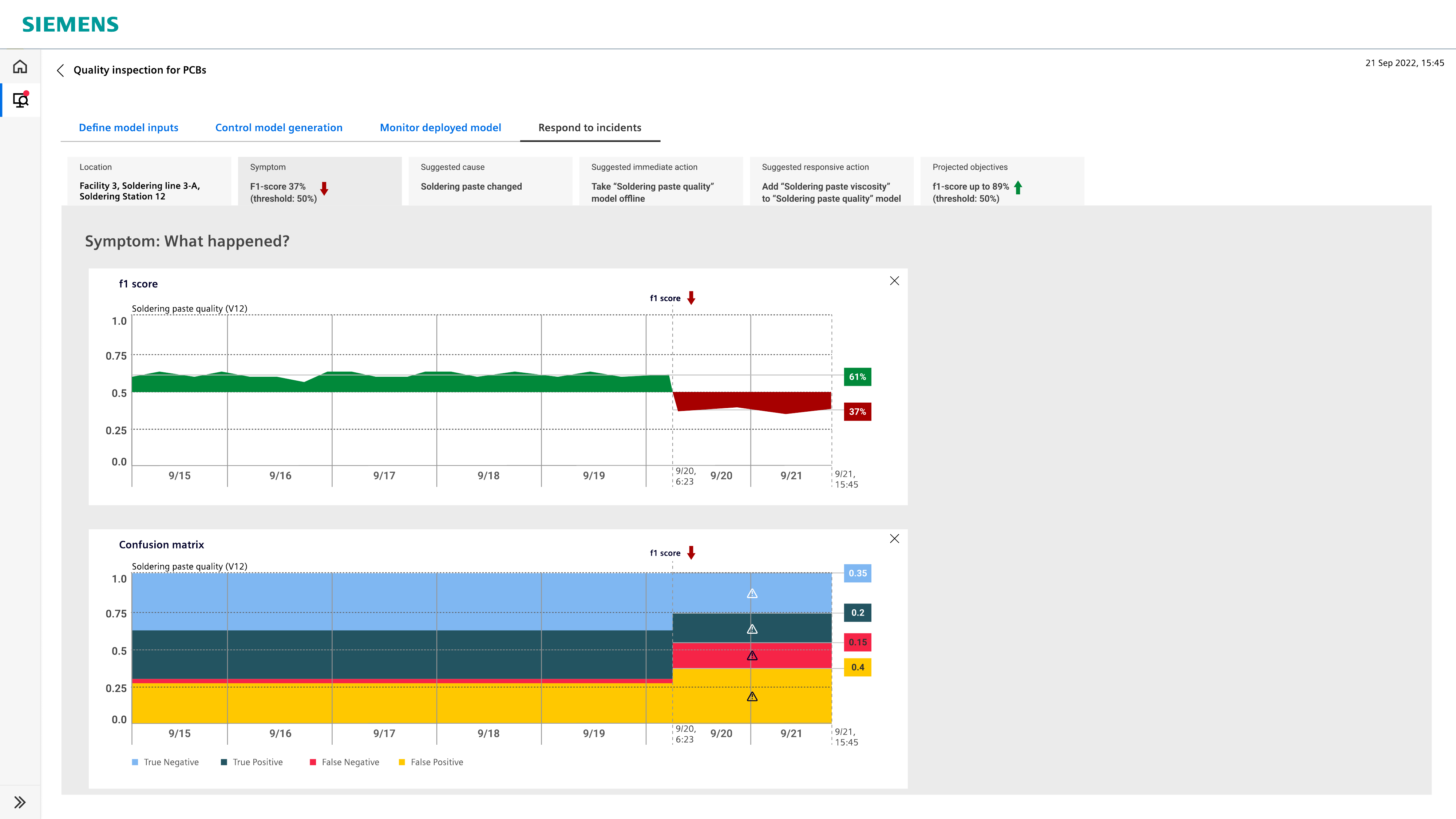}
	\caption{For user task "UT 3 Maintain ML-model"; select responsive action 1}
	\label{fig:Figma_Action1}
\end{figure}

\begin{figure}[h!]
	\includegraphics[width=1.5\textwidth,angle=90]{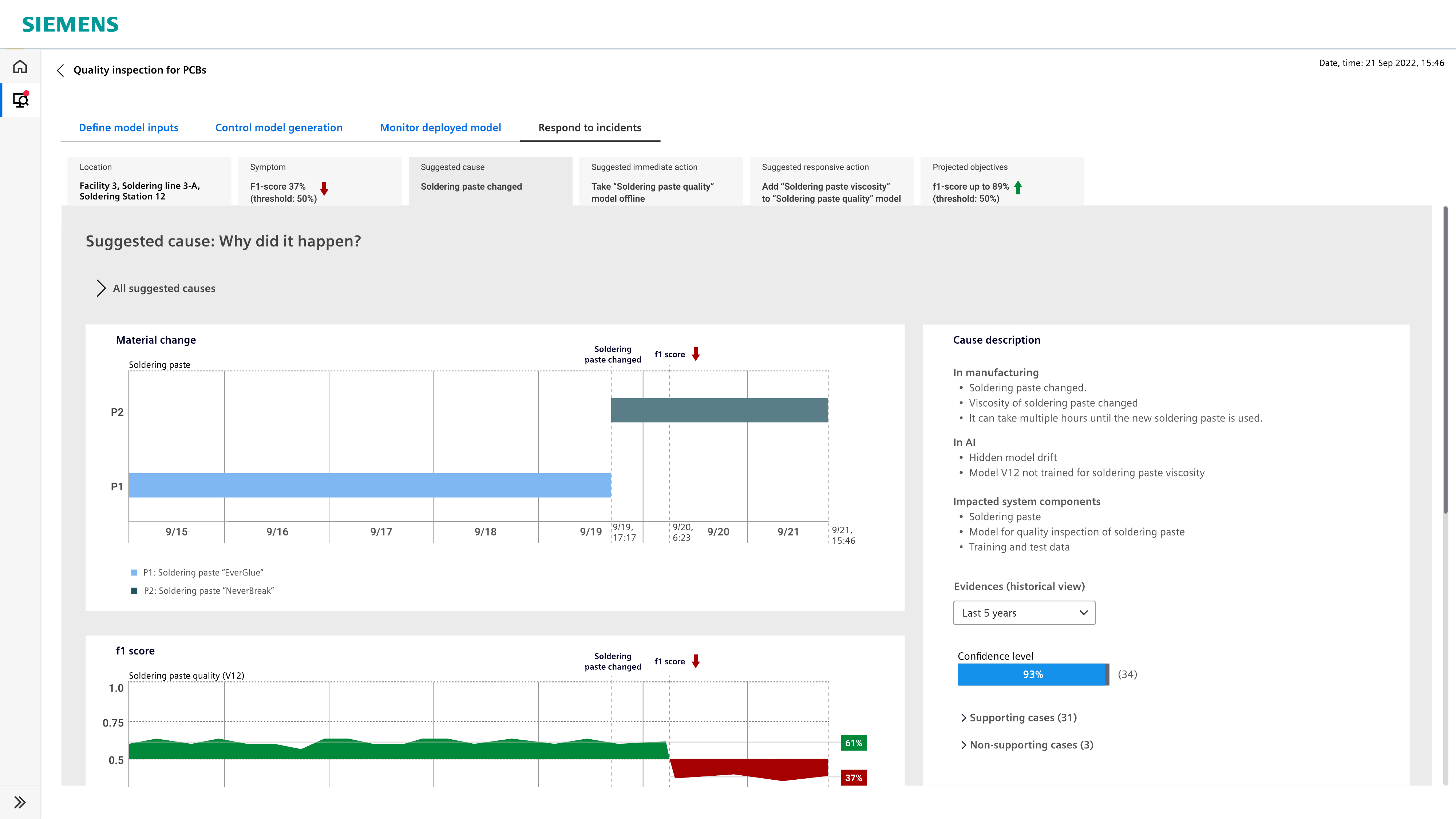}
	\caption{For user task "UT 3 Maintain ML-model"; select responsive action 2}
	\label{fig:Figma_Action2}
\end{figure}

\begin{figure}[h!]
	\includegraphics[width=1.5\textwidth,angle=90]{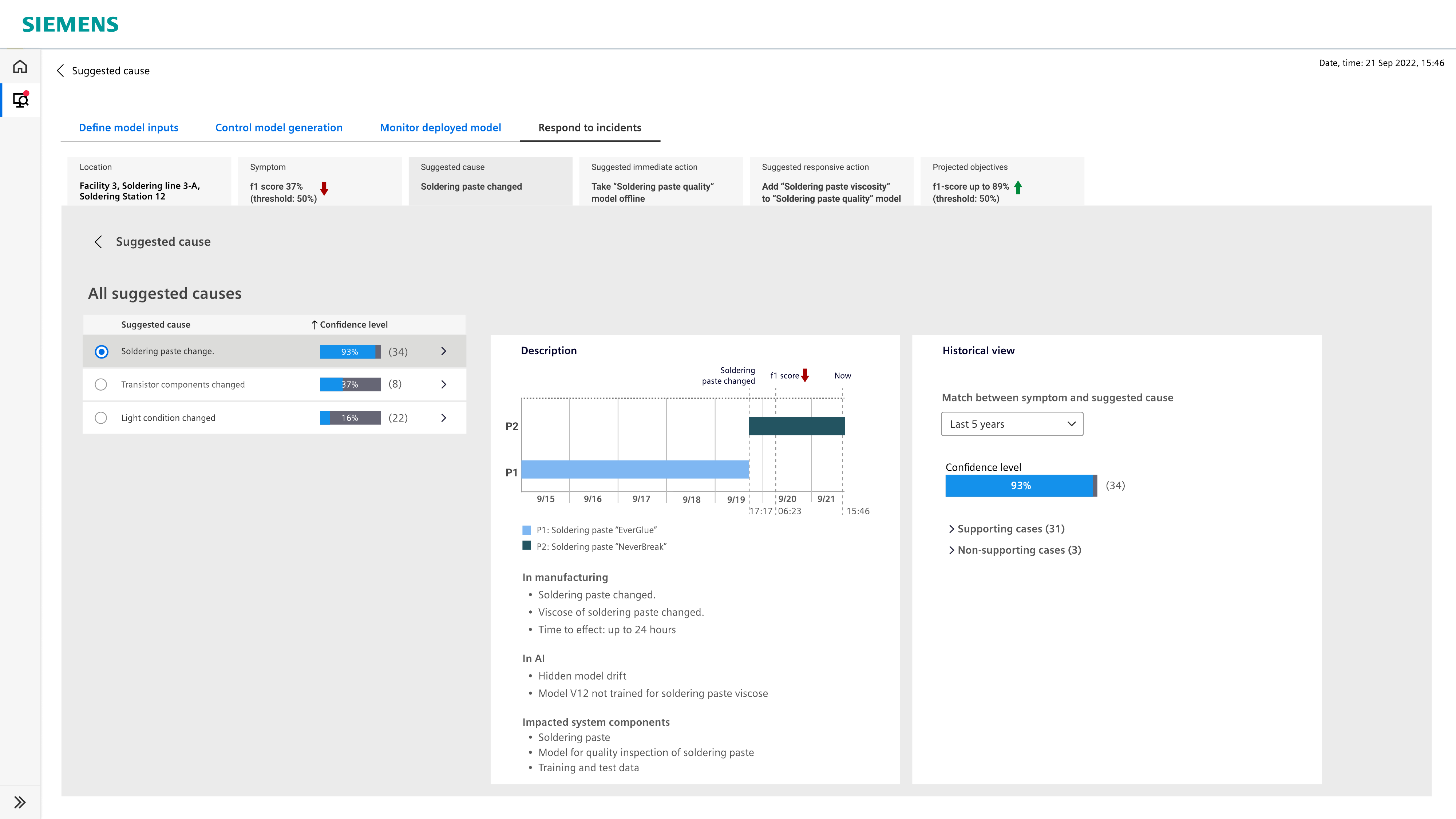}
	\caption{For user task "UT 3 Maintain ML-model"; select responsive action 3}
	\label{fig:Figma_Action3}
\end{figure}

\begin{figure}[h!]
	\includegraphics[width=1.5\textwidth,angle=90]{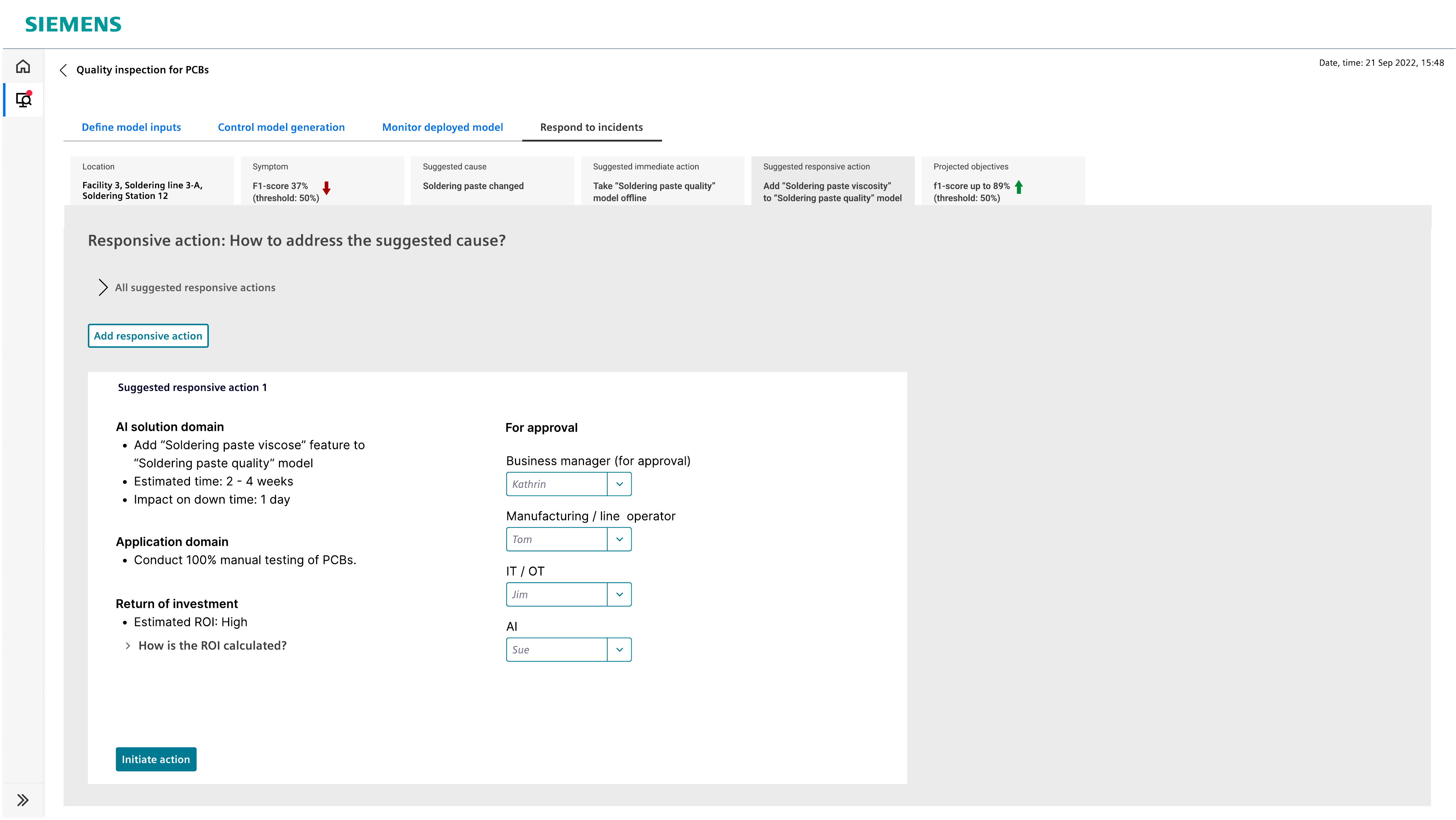}
	\caption{For user task "UT 3 Maintain ML-model"; select responsive action 4}
	\label{fig:Figma_Action4}
\end{figure}

\begin{figure}[h!]
	\includegraphics[width=1.5\textwidth,angle=90]{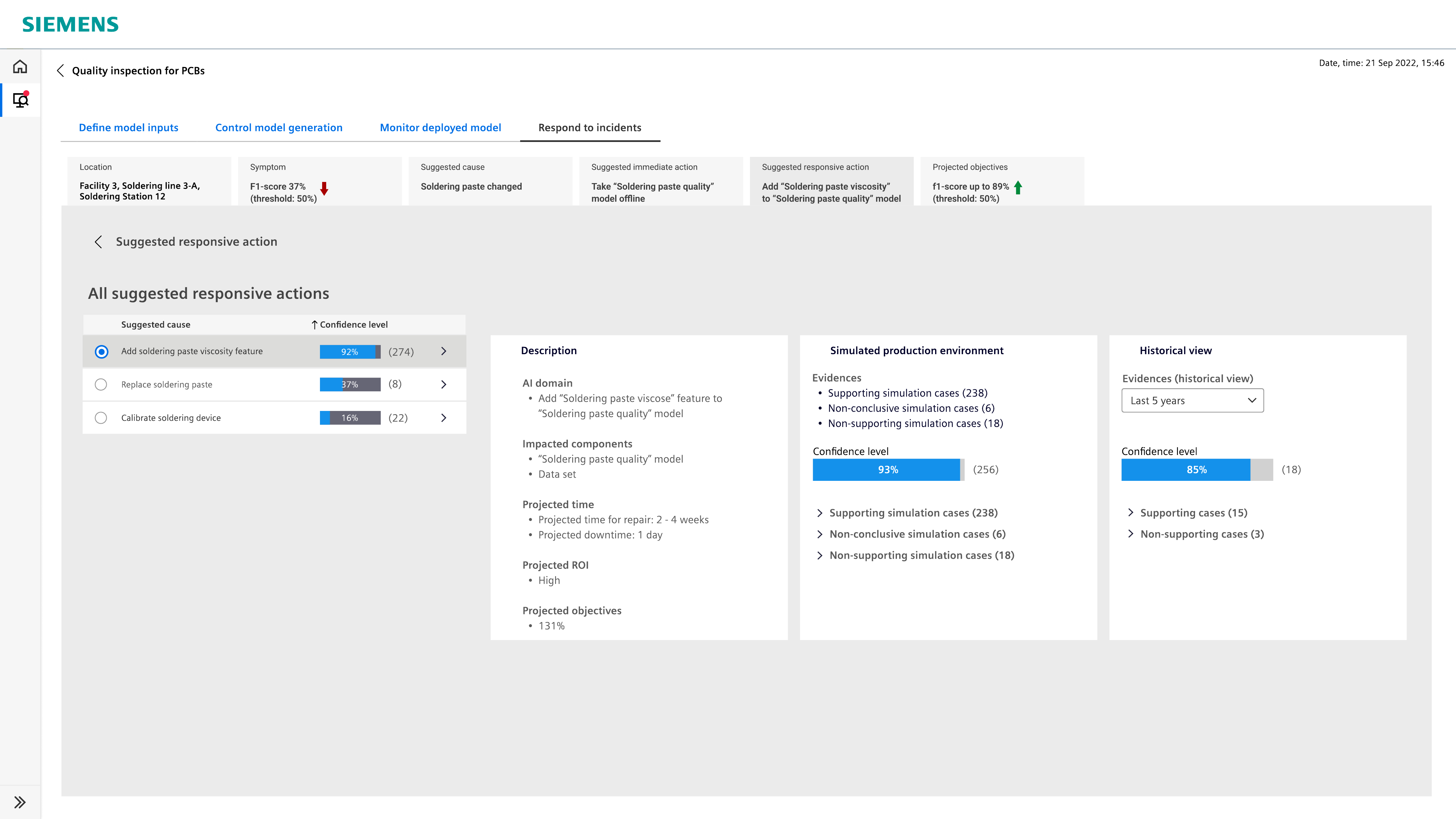}
	\caption{For user task "UT 3 Maintain ML-model"; select responsive action 5}
	\label{fig:Figma_Action5}
\end{figure}

\begin{figure}[h!]
	\includegraphics[width=1.5\textwidth,angle=90]{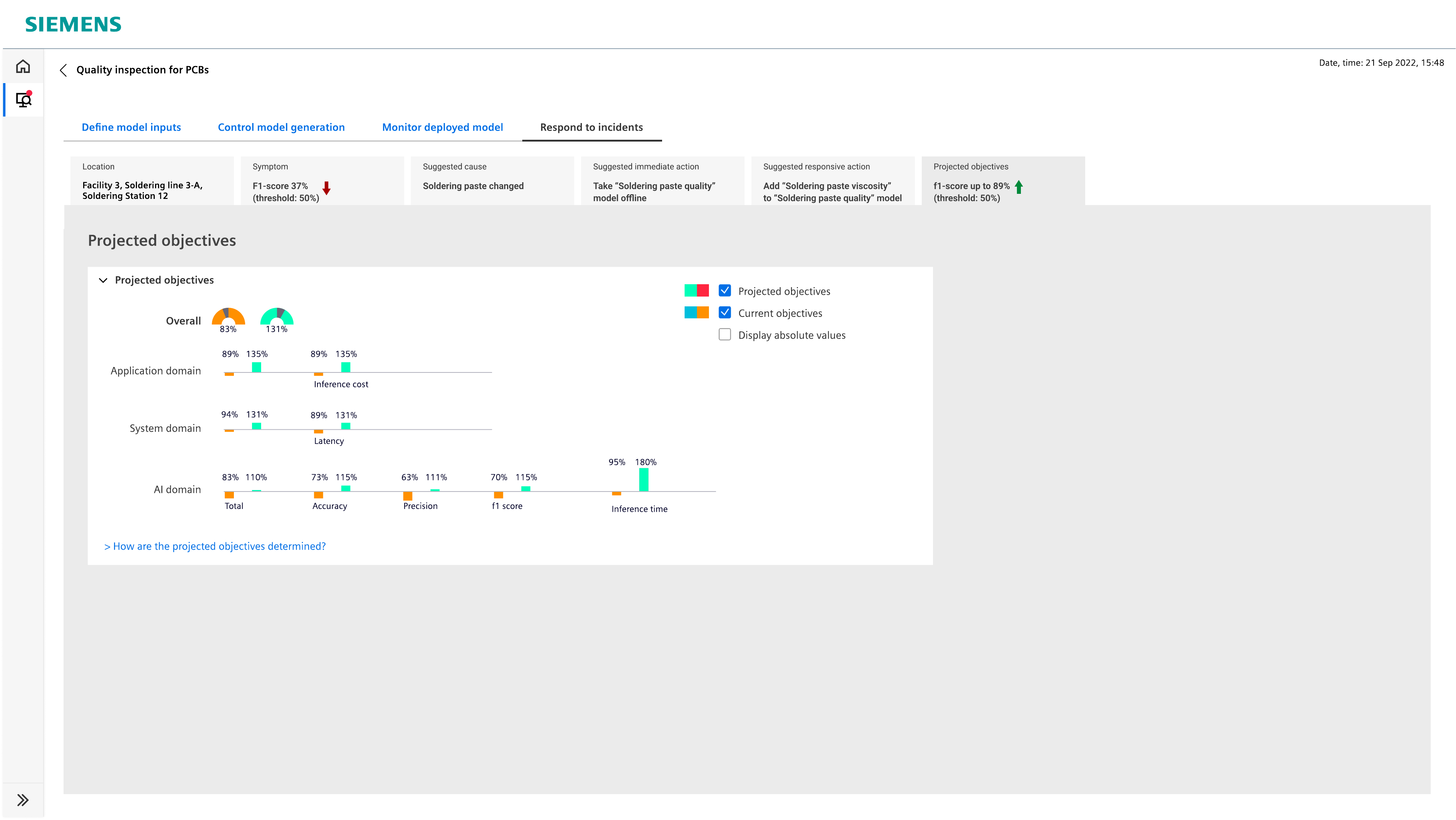}
	\caption{For user task "UT 3 Maintain ML-model"; select responsive action 6}
	\label{fig:Figma_Action6}
\end{figure}


\end{document}